\documentclass[twocolumn,english,showpacs,preprintnumbers,amsmath,amssymb,floatfix]{revtex4-1}
\usepackage{color}
\usepackage{array}
\usepackage{amstext}
\usepackage{amsmath}
\usepackage{graphicx}
\usepackage{esint}
\usepackage{rotating}
\usepackage[font=small,labelfont=bf]{caption}
\makeatletter

\@ifundefined{textcolor}{}
{%
 \definecolor{BLACK}{gray}{0}
 \definecolor{WHITE}{gray}{1}
 \definecolor{RED}{rgb}{1,0,0}
 \definecolor{GREEN}{rgb}{0,1,0}
 \definecolor{BLUE}{rgb}{0,0,1}
 \definecolor{CYAN}{cmyk}{1,0,0,0}
 \definecolor{MAGENTA}{cmyk}{0,1,0,0}
 \definecolor{YELLOW}{cmyk}{0,0,1,0}
 }

\@ifundefined{definecolor}
 {\usepackage{color}}{}
\@ifundefined{definecolor}
 {\usepackage{color}}{}
\makeatother

\makeatother

\usepackage{babel}
\begin{document}

\title{Fragmentation Functions of pion, kaon and proton at NLO approximation:  Laplace Transform approach}
\author{M.Zarei $^{a}$}
\email{m\_zarei\_128@yahoo.com}
\author{F.Taghavi-Shahri $^{a}$}
\email{f\_taghavi@ipm.ir}
\author{S. Atashbar Tehrani $^{b}$}
\email{atashbar@ipm.ir}
\author{M.Sarbishei $^{a}$}
\email{sarbishei@um.ac.ir}

\affiliation{
$^{(a)}$ Department of Physics, Ferdowsi University of Mashhad,
P.O. Box 1436, Mashhad, Iran\\
$^{(b)}$Independent researcher, P.O. Box 1149-8834413 Tehran, Iran \\  }

\date{\today}
\begin{abstract}
Using repeated Laplace transform, We find an analytical solution
for DGLAP evolution equations for extracting  the pion, kaon and
proton Fragmentation Functions (FFs) at NLO approximation. We also
study the symmetry breaking  of the sea quarks Fragmentation
Functions, $D_{\bar q}^h (z,Q^2)$ and simply separated them
according to their mass ratio.  Finally,  we calculate the total
Fragmentation Functions of these hadrons and compare them with
experimental data and those from global fits. Our results show a
good agreement with the FFs obtained from global parameterizations
as well as with the experimental data.
\end{abstract}


\maketitle

\section{Introduction}
Understanding the basic internal  structure of matter and the
quest for the ultimate constituents has always been important in
high energy physics. The nucleons are the basic building blocks of
atomic nuclei. The internal structure of the nucleons determines
their fundamental properties and directly affect the properties of
the nuclei. Therefore, understanding how the nucleon is built in
terms of its constituents
is an important and challenging question in modern nuclear physics.\\
Information about nucleon structure comes from two important
processes: The first one is semi- inclusive deep inelastic
scattering (SIDIS), whose reaction is as follows: $l+N \rightarrow
l+h+X$, and the second one is semi-inclusive hadron reaction like:
$p+p \rightarrow h+X$ . However, both of the processes require a
knowledge of the parton fragmentation
 functions (FFs) which describe the transition parton to hadron: $parton \rightarrow h+X$.\\ In general, fragmentation
  is the QCD process in which partons hadronize to colorless hadrons and the fragmentation
  functions, $D_i^h (z,Q^2)$,   represent the probability for a parton $i$ to fragments into
   a particular hadron $h$ carrying a certain fraction of the parton energy or momentum.
   They  are a necessary  ingredient in calculation of the single hadron inclusive production in any
    processes like $p\bar{p}$, $ep$, $\gamma p$ and $\gamma \gamma$ scattering.\\
Fragmentation functions cannot be computed directly from
perturbative QCD because,
 transition between color   partons into colorless hadrons is a soft/long-distance process, leading to
  divergences in the perturbation theory. Perturbative QCD dose not know anything about experimentally measured hadrons, but only quarks, anti-quarks
   and gluons. Fragmentation Functions can be evolved with DGLAP evolution equations from a starting distribution at a defined energy scale \cite{HALZEN,split}.

Recently we have used Laplace transform and provided an analytical
method to calculate Polarized Parton Distribution functions
(PPDFs)\cite{taghavi1,taghavi2}. In the present paper we will
apply this new method introduced by Block et
al.\cite{Block1,Block2,Block3,Block4,Block5,Block6} to calculate
pion, kaon and proton fragmentation functions. Therefore,  our
main task is finding analytical solutions of DGLAP evolution
equations to extract Fragmentation Functions (FFs). To do this, we
use the Laplace transform and find analytical solution of DGLAP
equations for FFs.  The initial inputs are selected from HKNS code
to warranty the correctness of our analytical calculations.
Finally, comparison of our FFs with those from global fits and
also with experimental data confirms the validity of our
calculations.

 The paper is organized as follows.
  In Section 2 we review  the method of analytical solution of DGLAP evolution equations for extracting Fragmentation Functions based on the Laplace transform.
  Then, in  Section 3  we utilize this method  to calculate   the Fragmentation Functions (FFs) of pion, kaon and proton.
  We also find a simple scenario for studying the symmetry breaking in the sea quarks FFs.
  Finally in section 4 we   calculated the total fragmentation   functions of pion, kaon and  proton and also compared them with available experimental
  data \cite{sld} and those from global fits \cite{hira1,hira2,kkp1,kkp2,DSS}.

\section{Analytical solution of DGLAP evolution equations for extracting Fragmentation Functions based on the Laplace transforms\label{sec:2}}
The Dokshitzer-Gribov-Lipatov-Altarelli-Parisi (DGLAP) evolution
equations \cite{DGLAP1,DGLAP2,DGLAP3}, for
the Fragmentation Functions (FFs) can be written as follows
\cite{split}:
\begin{eqnarray}
&&{4\pi\over \alpha_s(Q^2)}{\partial D_{ns}\over \partial \ln
(Q^2)}(z,Q^2)=D_{ns}\otimes\left[
P_{qq}^{LO,ns}\right.\nonumber \\
&&\left.+{\alpha_s(\tau)\over
4\pi}P_{qq}^{NLO,ns}\right](z,Q^2).\label{nonsingletinQsq_1}
\end{eqnarray}

\begin{eqnarray}
&&\frac{4\pi}{\alpha_s(Q^2)}\frac{\partial
D_s}{\partial\ln{Q^2}}(z,Q^2)= D_s\otimes \left(
P_{qq}^0+\frac{\alpha_s(Q^2)}{4\pi}P_{qq}^1\right)(z,Q^2)\nonumber \\
&&+D_g
\otimes \left(
P_{gq}^0+\frac{\alpha_s(Q^2)}{4\pi}P_{gq}^1\right)(z,Q^2),\label{DS}
\end{eqnarray}

\begin{eqnarray}
&&\frac{4\pi}{\alpha_s(Q^2)}\frac{\partial D_g }{\partial
\ln{Q^2}}(z,Q^2)= D_s\otimes \left(
P_{qg}^0+\frac{\alpha_s(Q^2)}{4\pi}P_{qg}^1\right)(z,Q^2)\nonumber \\
&&+D_g
\otimes \left(
P_{gg}^0+\frac{\alpha_s(Q^2)}{4\pi}P_{qg}^1\right)(z,Q^2).\label{DG}
\end{eqnarray}

where $P_{ij}^{0,1}$  are the leading and next to leading order
splitting functions. Block et al. in
Refs.\cite{Block1,Block2,Block3} showed that using the Laplace
transform, one can solve the DGLAP evolution equations directly
and extract unpolarized parton distribution functions . It is
possible to solve analytically the coupled leading and
next-to-leading-order DGLAP evolution equations to extract
Fragmentation Functions too. We will give the details here and
review the method for extracting the Fragmentation Functions at NLO approximation.\\
According to Block's scenario ,
 by introducing the variable $\nu\equiv
ln(\frac{1}{z})$ into the coupled DGLAP equations, one can turn
them into coupled convolution equations in $\nu$ space. Now, using
a new variable, namely, $\tau \equiv \frac{1}{4\pi
}\int_{Q_{0}^{2}}^{Q^{2}}\alpha _{s}(Q^{\prime 2})d\ln Q^{\prime
2}$, one can use two Laplace transforms from $\nu$ space to $s$
space and from $\tau$ space to $U$ space. With these two Laplace
transforms,  the DGLAP evolution equations can be solved
iteratively  by  a set of convolution integrals which are related
to Fragmentation Functions  at an initial input scale of $Q_0^2$.
Finally, two inverse Laplace transformations will back us to the
usual space ($z$, $Q^2$)\cite{taghavi2,Block6}.

\subsection{Non- Singlet  Fragmentation Functions}

At the NLO approximation, the fragmentation of valence quarks into
hadrons are given by  DGLAP evolution equations as:

\begin{eqnarray}
&&{4\pi\over \alpha_s(Q^2)}{\partial D_{ns}\over \partial \ln
(Q^2)}(z,Q^2)=D_{ns}\otimes\left[
P_{qq}^{LO,ns}\right.\nonumber \\
&&\left.+{\alpha_s(\tau)\over
4\pi}P_{qq}^{NLO,ns}\right](z,Q^2).\label{nonsingletinQsq_1}
\end{eqnarray}

where
 \begin{equation}
 D_{ns}^h (z,Q^2)= D_q^h
(z,Q^2) -  D_{\bar q}^h (z,Q^2)
\end{equation}

The  $\otimes$ symbol in the above equations refers to the
convolution integral in which the splitting functions in the
right-hand side of Eq. (4) are in fact functions of a variable
such as $\frac{x}{z}$.  Using the new  variables $\nu\equiv
ln(\frac{1}{z})$ , $w\equiv ln(\frac{1}{x})$ and $\tau \equiv
\frac{1}{4\pi }\int_{Q_{0}^{2}}^{Q^{2}}\alpha _{s}(Q^{\prime
2})d\ln Q^{\prime 2}$ and also defining  of  $z
D_{ns}(z,Q^2)=F_{ns}(z,Q^2)$, then we have the DGLAP evolution
equation as a function of  $\nu$ and $\tau$ variables as

\begin{eqnarray}
&&{\partial {\hat F_{ns}}\over \partial
\tau}(v,\tau)=\int_0^v{\hat F}_{ns}(w,\tau) e^{-(v-w)}\left[
P_{qq}^{LO,ns}(v-w)\right.\nonumber \\
&&\left.+{\alpha_s(\tau)\over
4\pi}P_{qq}^{NLO,ns}(v-w)\right]d\,w.\label{nonsingletinQsq_2}
\end{eqnarray}

where
\begin{equation}
{\hat F}_{ns}(v,\tau )\equiv
F_{ns}(e^{-v},\tau)),\label{FandGhat2}
\end{equation}

Because the r.h.s of Eq. (6) is a normal convolution integral, we
 can use the following property for the product of Laplace
transform  :

\begin{equation}
{\cal L}\left[\int_0^v {\hat F}[w]{\hat H}[v-w]\,dw;s
\right]={\cal L} [{\hat F}[v];s]\times {\cal L} [{\hat
H}[v];s]\label{convolution}.
\end{equation}

Then we will get a simple solution for valence fragmentation
functions in $s$ space:

\begin{equation}
f_{ns}(s,\tau)=e^{\tau
\Phi_{ns}(s)}f_{ns0}(s),\label{solutionandPhinsofs}
\end{equation}

in which
\begin{equation}
\Phi_{ns}(s)\equiv \Phi_{ns}^{LO}(s)+{\tau_2\over
\tau}\Phi_{ns}^{NLO}(s),
\end{equation}
where
\begin{eqnarray}
&&\Phi_{ns}^{LO}(s)\equiv{\cal L}\left
[e^{-v}P_{qq}^{LO,ns}(e^{-v});s\right
],\nonumber\\
&&\Phi_{ns}^{NLO}(s)\equiv{\cal L}\left
[e^{-v}P_{qq}^{NLO,ns}(e^{-v});s\right ].\label{PhiNs0and1}
\end{eqnarray}

The Laplace transform of non- singlet splitting functions,
$\Phi_{ns}^{LO}(s)$ and $\Phi_{ns}^{NLO}(s)$ are given in
Appendix. A. The  $\tau_2$ parameter
 in Eq. (10) is defined as

\begin{eqnarray}
\tau_2&\equiv& {1\over 4\pi}\int_0^\tau\alpha_s(\tau')\,d\tau'
={1\over (4\pi)^2}\int_{Q^2_0}^{Q^2}\alpha_s^2(Q'^2)\, d \ln
Q'^2,\label{tau2ofQsq}\nonumber\\
\end{eqnarray}

 In Eq.(9) the $f_{ns0}(s)$ function is the Laplace transform of
valence quark fragmentation functions at initial scale of
$Q_0^2=4.5 GeV^2$. We got them from HKNS code \cite{hira1}.
Finally employing the inverse Laplace transform on Eq.
(9)\cite{Block6}, we can derive the valence quark fragmentation
functions in $(z,Q^2)$ space.

\begin{figure}
\includegraphics[clip,width=0.4\textwidth]{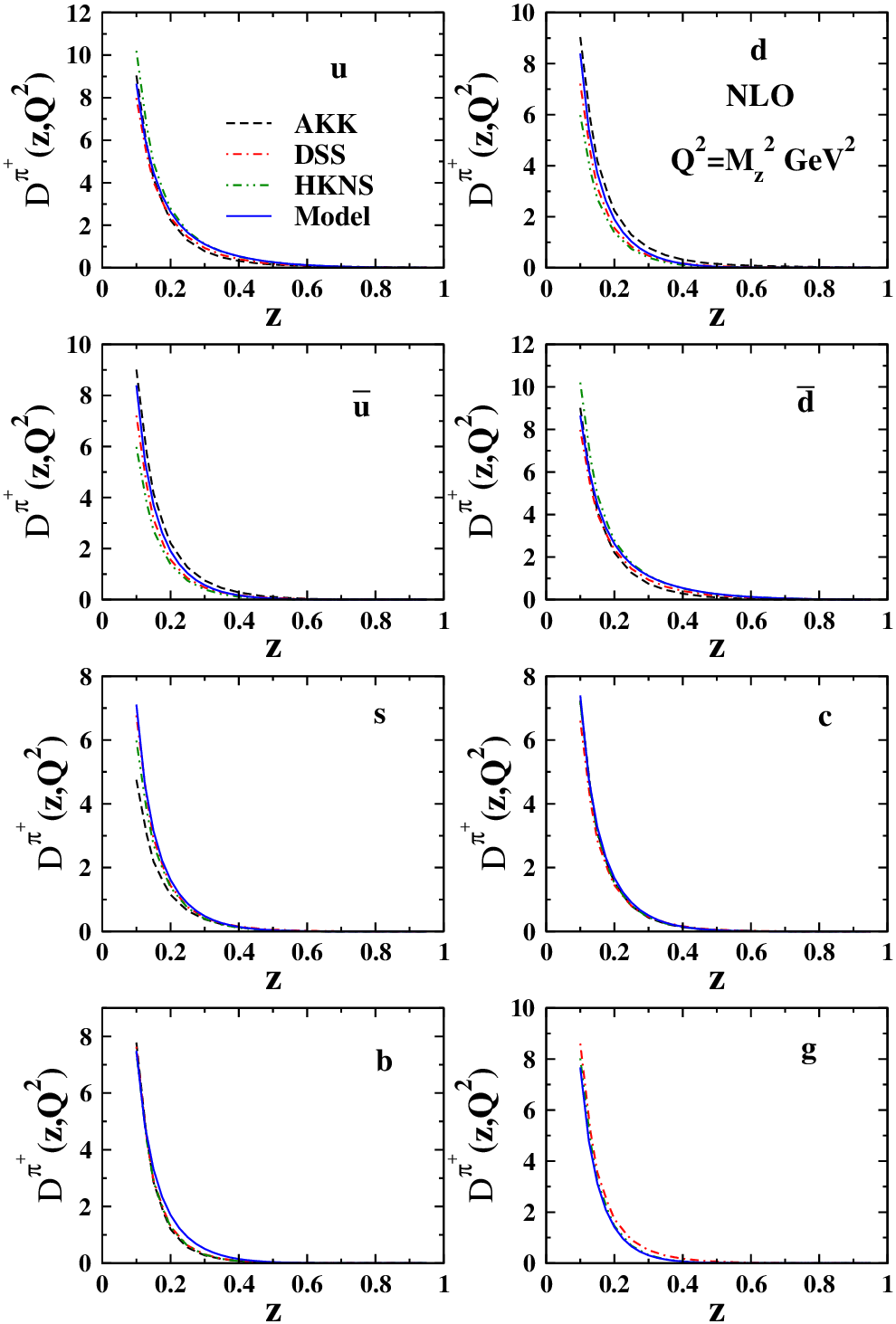}
\begin{center}
\caption{Pion fragmentation functions at $Q^2=M_z^2$ and
Comparison with AKK, DSS and  HKNS  global fits.}\label{fig:figQ0}
\end{center}
\end{figure}

\begin{figure}
\includegraphics[clip,width=0.4\textwidth]{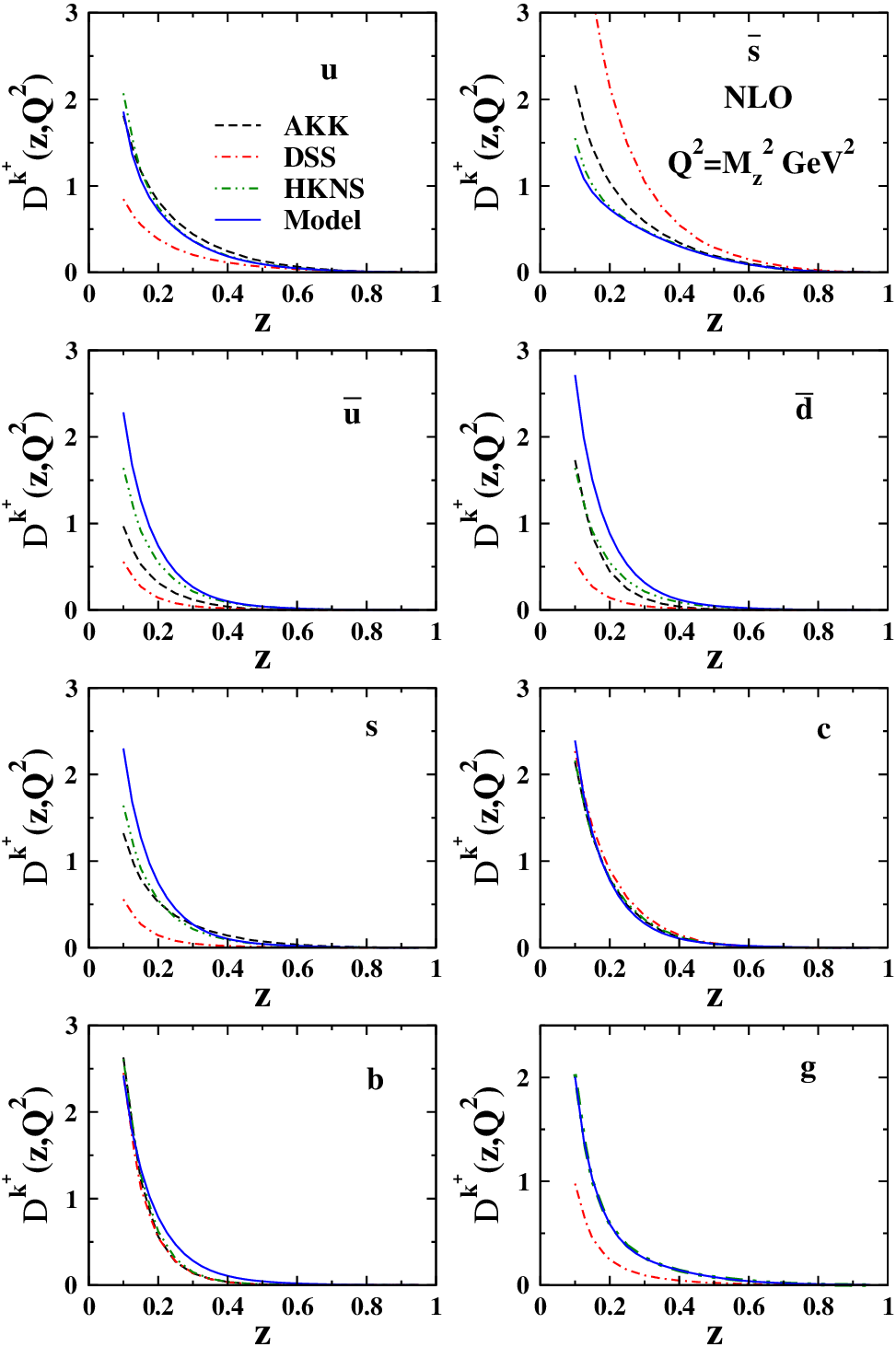}
\begin{center}
\caption{Kaon fragmentation functions at $Q^2=M_z^2$and
Comparison with AKK, DSS and  HKNS  global fits.}\label{fig:figQ0}
\end{center}
\end{figure}

 \begin{figure}
\includegraphics[clip,width=0.4\textwidth]{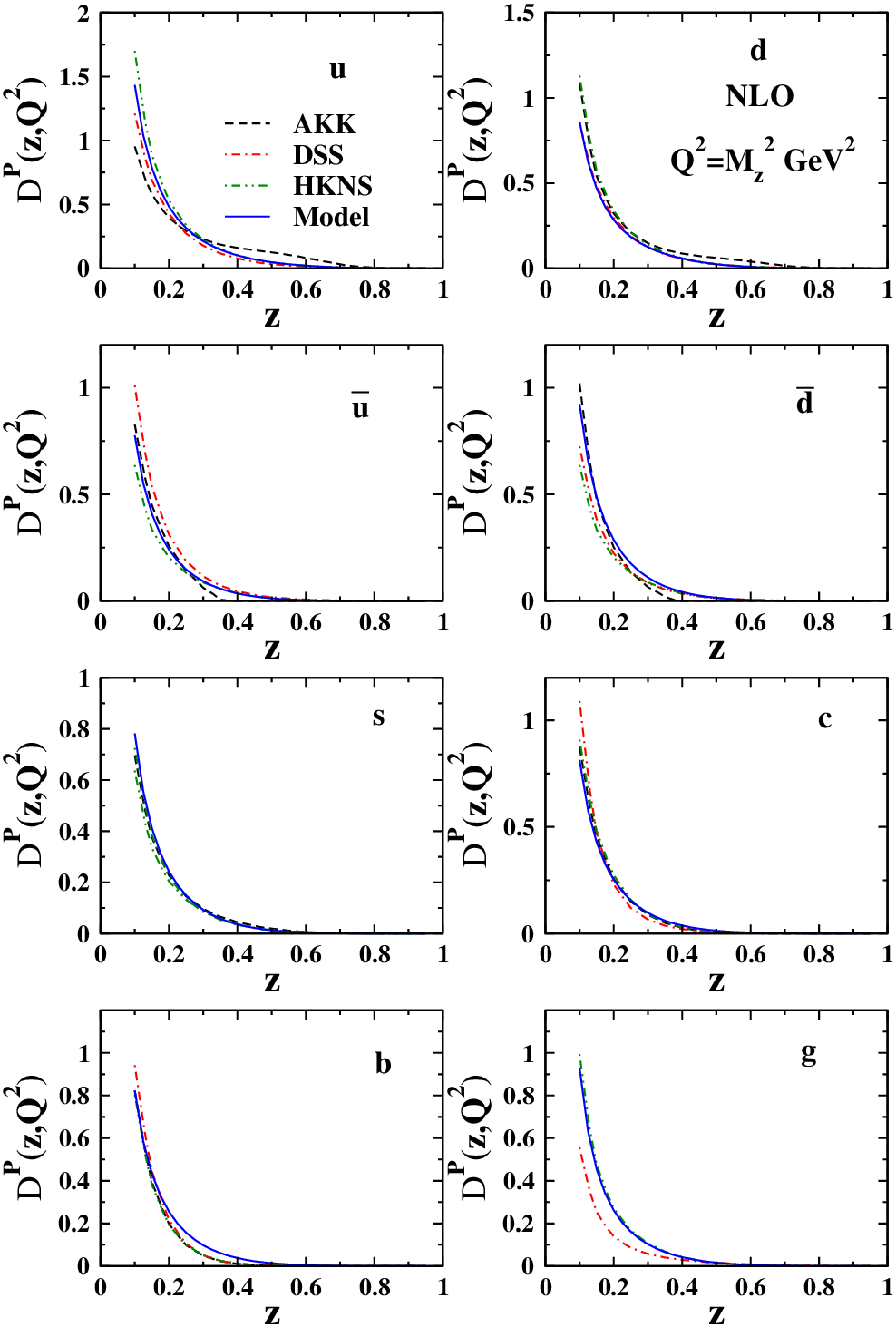}
\begin{center}
\caption{Proton fragmentation functions at $Q^2=M_z^2$and
Comparison with AKK, DSS and  HKNS  global fits.}\label{fig:figQ0}
\end{center}
\end{figure}

\begin{figure}
\includegraphics[clip,width=0.3\textwidth]{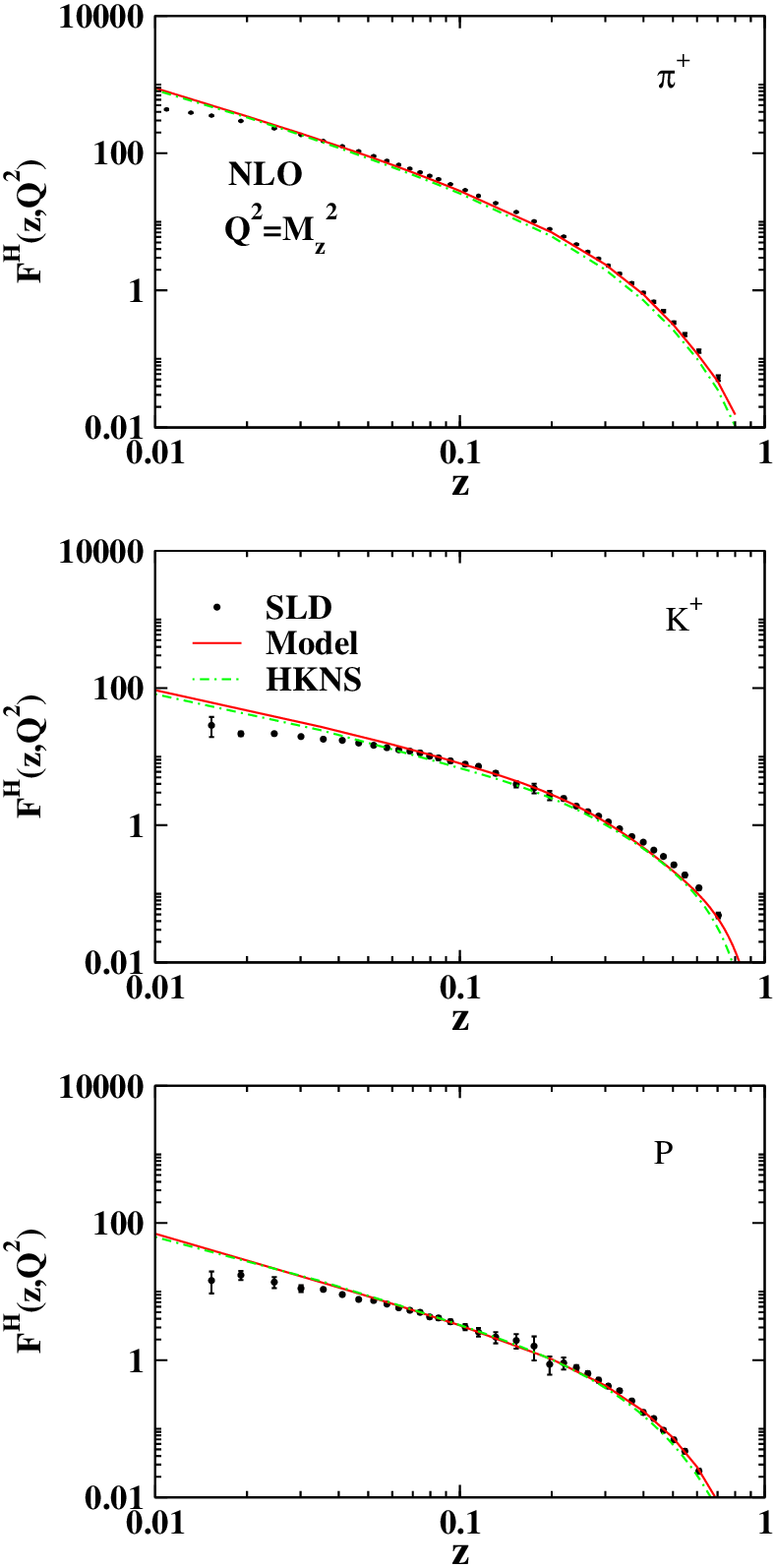}
\begin{center}
\caption{Total fragmentation functions of pion, kaon and proton
and comparison with experimental data from SLD \cite{sld} at
$Q^2=M_z^2$. We also compared our results with
 HKNS global fit.}\label{fig:figQ0}
\end{center}
\end{figure}

\subsection{Singlet and gluon Fragmentation Functions}
The coupled NLO DGLAP evolution equations for extracting the
singlet and gluon fragmentation functions are given as follows
\begin{eqnarray}
&&\frac{4\pi}{\alpha_s(Q^2)}\frac{\partial
D_s}{\partial\ln{Q^2}}(z,Q^2)= D_s\otimes \left(
P_{qq}^0+\frac{\alpha_s(Q^2)}{4\pi}P_{qq}^1\right)(z,Q^2)\nonumber\\
&&+D_g\otimes \left(
P_{gq}^0+\frac{\alpha_s(Q^2)}{4\pi}P_{gq}^1\right)(z,Q^2),\label{DS}
\end{eqnarray}

\begin{eqnarray}
&&\frac{4\pi}{\alpha_s(Q^2)}\frac{\partial D_g }{\partial
\ln{Q^2}}(z,Q^2)= D_s\otimes \left(
P_{qg}^0+\frac{\alpha_s(Q^2)}{4\pi}P_{qg}^1\right)(z,Q^2)\nonumber\\
&&+D_g\otimes \left(
P_{gg}^0+\frac{\alpha_s(Q^2)}{4\pi}P_{qg}^1\right)(z,Q^2).\label{DG}
\end{eqnarray}

where the singlet fragmentation function is defined as

\begin{equation}
D_s^h (z,Q^2)=\sum_{q={u,d,s,c,b}} [ D_q^h (z,Q^2) +  D_{\bar q}^h
(z,Q^2) ]
\end{equation}

Using the convention  $z D_{s}(z,Q^2) \equiv F_{s}(z,Q^2)$ and $z
D_{g}(z,Q^2) \equiv G(z,Q^2)$, these coupled equations can be
written  in terms of $\nu$ and $\tau$ variables, which have been
defined in previous section. Then we will arrive at:

\begin{eqnarray}
&&\frac{\partial {\hat F}_s}{\partial \tau}(v,\tau)= \int_0^v {\hat
F}_s(w,\tau) \left({\hat H}_{qq}(v-w)\right.\nonumber\\
&&\left.+\frac{\alpha_s(\tau)}{4\pi}{\hat H}_{qq}^1(v-w)\right)d\,w
\nonumber \\
&&+\int_0^v {\hat G}(w,\tau) \left({\hat H}_{gq}(v-w)\right.\nonumber\\
&&\left.+\frac{\alpha_s(\tau)}{4\pi}{\hat H}_{gq}^1(v-w)\right)d\,w,
\label{dsvwt}
\end{eqnarray}
\begin{eqnarray}
&&\frac{\partial {\hat G}}{\partial \tau}(v,\tau)=\int_0^v {\hat
F}_s(w,\tau) \left({\hat H}_{qg}(v-w)\right.\nonumber\\
&&\left.+\frac{\alpha_s(\tau)}{4\pi}{\hat H}_{qg}^1(v-w)\right)d\,w
\nonumber\\
&&+\int_0^v {\hat G}(w,\tau) \left({\hat H}_{gg}(v-w)\right.\nonumber\\
&&\left.+\frac{\alpha_s(\tau)}{4\pi}{\hat
H}_{gg}^1(v-w)\right)d\,w,
\label{dgvwt}
\end{eqnarray}

in which we use the definitions:

\begin{equation}
{\hat H}_{ij}^0(v)\equiv  e^{-v}P_{ij}^0(e^{-v}), {\hat
H}_{ij}^1(v)\equiv  e^{-v}P_{ij}^1(e^{-v}),\label{HNLO}
\end{equation}

\begin{equation}
{\hat F}_s(v,\tau )\equiv  F_s(e^{-v},\tau),\qquad \hat G(v,\tau
)\equiv  G(e^{-v},\tau),\label{FandGhat2}
\end{equation}

At NLO approximation we need two Laplace transforms to decouple
the DGLAP equations into two simple equations that can be solved
iteratively. The first Laplace transform  from  $\nu$ space to $s$
space changes the DGLAP evolution equation to the first order
coupled differential equations as

\begin{eqnarray}
&&{\partial f\over \partial \tau }(s,\tau) =\left(
\Phi_f^{LO}(s)+\frac{\alpha_s(\tau )}{4\pi}
\Phi_f^{NLO}(s)\right)f(s,\tau)\nonumber\\
&&+\left(\Theta_g^{LO}(s)+\frac{\alpha_s(\tau)}{4\pi}\Theta_g^{NLO}(s)\right)g(s,\tau)\label{df},
\end{eqnarray}

\begin{eqnarray}
&&{\partial g\over \partial \tau }(s,\tau) =\left( \Phi_
g^{LO}(s)+\frac{\alpha_s(\tau)}{4\pi}
\Phi_g^{NLO}(s)\right)g(s,\tau)\nonumber\\
&&+\left(\Theta_f^{LO}(s)+\frac{\alpha_s(\tau)}{4\pi}\Theta_f^{NLO}(s)\right)f(s,\tau),\label{dg}
\end{eqnarray}

where

\begin{equation}
f(s,\tau) \equiv {\cal L}[ \hat F_s(v,\tau);s],\qquad
g(s,\tau)\equiv {\cal L}[\hat G(v,\tau);s]
\end{equation}

The Laplace transform of  singlet and gluon splitting functions,
$\Phi_{f,g}^{LO,NLO}(s)$ are given in Appendix. A. The second
Laplace transform from $\tau$ space to $U$ space changes  Eq. (20)
and Eq. (21) into two simple linear algebraic equations as

\begin{eqnarray}
&&U{\cal F}(s,U)-f_0(s)=\nonumber\\
&&\Phi_f^{LO}(s){\cal
F}(s,U)+\Phi_f^{NLO}(s){\cal L}[\frac{\alpha_s(\tau
)}{4\pi} f(s,\tau);U]\nonumber\\
&&+\Theta_g^{LO}(s){\cal G}(s,U)+\Theta_g^{NLO}(s){\cal
L}[\frac{\alpha_s(\tau)}{4\pi}g(s,\tau);U],\label{u,ta}
\end{eqnarray}

\begin{eqnarray}
&&U{\cal G}(s,U)-g_0(s)=\nonumber\\
&&\Phi_g^{LO}(s){\cal
G}(s,U)+\Phi_g^{NLO}(s){\cal L}[\frac{\alpha_s(\tau
)}{4\pi} g(s,\tau);U]\nonumber\\
&&+\Theta_f^{LO}(s){\cal F}(s,U)+\Theta_f^{NLO}(s){\cal
L}[\frac{\alpha_s(\tau)}{4\pi}f(s,\tau);U].\label{u,ta,g}
\end{eqnarray}

where we have

\begin{eqnarray}
{\cal F}(s,U)&\equiv &{\cal L}\left[f(s,\tau);U\right ],\qquad
{\cal G}(s,U)\equiv {\cal L}\left[g(s,\tau);U\right ],\nonumber\\\label{LAP
U}
\end{eqnarray}

\begin{eqnarray}
{\cal L}\left[{\partial f\over \partial \tau }(s,\tau);U\right]&=&U{\cal F}(s,U)-f_0(s),\nonumber\\
\qquad {\cal L}\left[{\partial g\over \partial \tau
}(s,\tau);U\right]&=&U{\cal G}(s,U)-g_0(s),\label{D LAPG}
\end{eqnarray}

 To simplify the NLO calculations we use an excellent
approximation relation
 $a(\tau)={\alpha_s(\tau)\over 4\pi} \approx
a_0+ a_1e^{-b_1\tau}$, where $a0 = 0.0037, a1 = 0.025, b1 = 10.7$
 \cite{Block1}. Therefore we
write the Laplace transform of ${\cal L}
[\frac{\alpha_s(\tau)}{4\pi} f(s,\tau);U]$ and ${\cal
L}[\frac{\alpha_s(\tau)}{4\pi} g(s,\tau);U]$ which are needed in
Eq. (24) and Eq. (25) as

\begin{eqnarray}
{\cal L}[\frac{\alpha_s(\tau)}{4\pi} f(s,\tau);U]&=&\sum_{j=0}^1a_j{\cal F}(s,U+b_j),\qquad \nonumber\\
{\cal L}[\frac{\alpha_s(\tau)}{4\pi} g(s,\tau);U]&=&\sum_{j=0}^1
a_j{\cal G}(s,U+b_j), b_0=0\label{transformoffalpha}
\end{eqnarray}

Now we define

\begin{eqnarray}
&&\Phi_f(s)\equiv \Phi_f^{LO}(s)+a_0\Phi_f^{NLO}(s),\nonumber\\
&&\Phi_g(s)\equiv \Phi_g^{LO}(s)+a_0\Phi_g^{NLO}(s),\label{Phis}
\end{eqnarray}

\begin{eqnarray}
&&\Theta_f(s)\equiv \Theta_f^{LO}(s)+a_0\Theta_f^{NLO}(s),\nonumber\\
&&\Theta_g(s)\equiv
\Theta_g^{LO}(s)+a_0\Theta_g^{NLO}(s),\label{Thetas}
\end{eqnarray}

Finally, in  $s$ and $U$ space, we arrive at the following two coupled algebraic
equations for singlet and gluon fragmentation functions which can
be solved by iteration method described in \cite{taghavi2,Block1}:

\begin{eqnarray}
&&\left[ U -\Phi_f(s)\right]{\cal F}(s,U)- \Theta_g(s){\cal G}(s,U)= f_0(s)\nonumber\\
&&+ a_1\left[\Phi_f^{NLO}(s){\cal F}(s,U+b_1)+\Theta_g^{NLO}(s)
{\cal G}(s,U+b_1)\right] ,\nonumber\\
\label{Feqn}
\end{eqnarray}

\begin{eqnarray}
&&-\Theta_f(s){\cal F}(s,U)+\left[ U -\Phi_g(s)\right]{\cal G}(s,U)= g_0(s)\nonumber\\
&&+ a_1\left[\Theta_f^{NLO}(s){\cal
F}(s,U+b_1)+\Phi_g^{NLO}(s){\cal G}(s,U+b_1)\right] ,\nonumber \\
\label{Geqn}
\end{eqnarray}

With the initial input functions for singlet (sum of valence and
sea quarks) and gluon sectors of distributions, which are denoted
by $f_0(s)$ and $g_0(s)$ respectively, their evolved solutions in
the Laplace s space are given by \cite{Block5}

\begin{eqnarray}
f(s,\tau)=k_{ff}(s,\tau)f_0(s)+ k_{fg}(s,\tau)g_0(s)\nonumber\\
g(s,\tau)=k_{gg}(s,\tau)g_0(s)+ k_{gf}(s,\tau)f_0(s)
\end{eqnarray}

where the $ k$'s  in Eq. (32) have been
introduced in Refs. \cite{Block1}. These function are given in
Appendix. B for the first iteration. The initial inputs are
selected  from HKNS code \cite{hira1} at initial scale of
$Q_0^2=4.5 GeV^2$.
 Finally with an
inverse laplace transform  one can derive the singlet and gluon
fragmentation functions in $(z,Q^2)$ space \cite{Block6}. It
should be noted that our initial inputs are quoted from HKNS code
to confirm the validity of our analytical solutions. If we reach
to the acceptable agreement between our FFs and FFs obtained by
global fits and also with  those from experimental data,  then we
can be  sure that our analytical solution for FFs are correct. In
the next work, this method is employed to yield us the initial
inputs via global fit to experimental data.

\section{Pion, Kaon and Proton Fragmentation Functions }

In this section, we present the results of partons fragmentation
functions of pion, kaon and proton. As we did in the last
sections, we can calculate the non-singlet, singlet and gluon
Fragmentation Functions using analytical solution of DGLAP
evolution equations in Laplace space $(s,\tau)$. Then, with an
inverse laplace transform  the valence ,  singlet and gluon
Fragmentation Functions in $(z,Q^2)$ space are obtained. In this
connection we need to use the flavor symmetries between different
kinds of fragmentation functions in pion, kaon or proton at scale
of $Q^2$ as it follows: \cite{hira1}:

\begin{eqnarray}
D_{\bar u}^{\pi^+} (z,Q^2)  = D_{d}^{\pi^+} (z,Q^2)
     \ne D_{s}^{\pi^+} (z,Q^2)\nonumber\\
 D_{u}^{\pi^+} (z,Q^2) = D_{\bar d}^{\pi^+} (z,Q^2)\nonumber\\
 D_{s}^{\pi^+} (z,Q^2) = D_{\bar s}^{\pi^+} (z,Q^2)\nonumber\\
 D_{c}^{\pi^+} (z,Q^2) = D_{\bar c}^{\pi^+} (z,Q^2)\nonumber\\
 D_{b}^{\pi^+} (z,Q^2) = D_{\bar b}^{\pi^+} (z,Q^2)
\end{eqnarray}
\begin{eqnarray}
D_{\bar u}^{K^+} (z,Q^2) \ne D_{d}^{K^+} (z,Q^2)
                         \ne D_{s}^{K^+} (z,Q^2)\nonumber\\
 D_{d}^{K^+} (z,Q^2) = D_{\bar d}^{K^+} (z,Q^2)\nonumber\\
 D_{c}^{K^+} (z,Q^2) = D_{\bar c}^{K^+} (z,Q^2)\nonumber\\
 D_{b}^{K^+} (z,Q^2) = D_{\bar b}^{K^+} (z,Q^2)
\end{eqnarray}

\begin{eqnarray}
D_{u}^{p^+}(z,Q^2)  \ne 2 \, D_{d}^{p^+} (z,Q^2)\nonumber\\
D_{\bar u}^{p^+}(z,Q^2) \ne D_{\bar d}^{p^+} (z,Q^2)\ne D_{s}^{p^+} (z,Q^2)\nonumber\\
D_{s}^{p^+}(z,Q^2)= D_{\bar s}^{p^+} (z,Q^2)\nonumber\\
D_{c}^{p^+}(z,Q^2)= D_{\bar c}^{p^+} (z,Q^2)\nonumber\\
D_{b}^{p^+}(z,Q^2)= D_{\bar b}^{p^+} (z,Q^2)
\end{eqnarray}

\subsection{Symmetry breaking in the sea quarks Fragmentation Functions}

The total sea quarks fragmentation function is calculated as follows

\begin{eqnarray}
 D_{s}(z,Q^2)- D_{ns}(z,Q^2)= D_{\bar q}(z,Q^2)
\end{eqnarray}
Where $ D_{\bar q}(z,Q^2)$ is
\begin{eqnarray}
&& D_{\bar q}(z,Q^2)=2 D_{\bar u}(z,Q^2)+2 D_{\bar d}(z,Q^2)+2 D_{s}(z,Q^2)\nonumber\\
 &&+2 D_{c}(z,Q^2)+2 D_{b}(z,Q^2),
\end{eqnarray}
  Now to investigate  the symmetry breaking of sea quarks
 fragmentation functions we use the fact that
    heavier sea quarks can produce hadrons with higher
  probability. Therefore, the fraction of  different kind of sea
  quarks can be proportional to their mass ratio.
 For  example  we have  $\frac{D_{\bar
 u}}{D_{c}}\simeq\frac{m_u}{m_c}$. As an example, if we want to
 calculate the c quark fragmentation function, we have

\begin{eqnarray}
&&D_{\bar q}(z,Q^2)=2 \frac{m_u}{m_c} D_{c}(z,Q^2)+2 \frac{m_d}{m_c} D_{c}(z,Q^2)\nonumber \\
&+&2 \frac{m_s}{m_c} D_{c}(z,Q^2)+2 D_{c}(z,Q^2)+2\frac{m_b}{m_c}
D_{c}(z,Q^2),
\end{eqnarray}

\begin{eqnarray}
D_{c}(z,Q^2)\simeq\frac{D_{\bar q}(z,Q^2)}{(2 \frac{m_u}{m_c}+2
\frac{m_d}{m_c} +2 \frac{m_s}{m_c}+2 +2\frac{m_b}{m_c}) }
\end{eqnarray}
This leads to the following general relation:

\begin{eqnarray}
 D_{quark} (z,Q^2)=\frac{D_{\bar q}(z,Q^2)}{B^A},
\end{eqnarray}

where B is the mass ratio and it is constant parameter for each
kind of sea quark. The free parameter A should be extracted from
experimental data,  however we have  used HKNS code for extracting
the sea quarks FFs to be sure about our analytical solutions. The
results are listed  in Table 1 . The results for all fragmentation
functions for pion, kaon and proton at $Q^2=M_z^2$ are shown in
figures 1, 2 and 3 respectively. We also compared our FFs with
those from global fits of HKNS, AKK and DSS groups
\cite{hira1,hira2,kkp1,kkp2,DSS}. They is good agreement between
them. The results show that our analytical solutions for DGLAP
evolution equations are correct and these solutions are correctly
used to calculate the Fragmentation Functions. In the next section
we will calculate total Fragmentation Functions of pion, kaon and
proton to test our calculations with experimental data.

\begin{table}
 {\footnotesize
\centerline{\begin{tabular}{|c|c|c|c|c}
  \hline
   $quarks$  & $A$ & $B$\\
     \hline
$\bar u $ &0.25& 4651 \\
  \hline
  $\bar d $& 0.25 & 2325.5\\
   \hline
  $s $&0.45 &107.33\\
   \hline
  $c $  & 0.95  & 8.7893\\
   \hline
   $b$   & 2.1  & 2.6577 \\
   \hline
\end{tabular}}
 \caption{\label{a2}  Parameters A and B  in the sea quark fragmentation functions. }}
\end{table}

\section{ Total Fragmentation Functions of pion, kaon and proton}

In this section we intend to calculate the total hadron
fragmentation function for pion, kaon and proton. We use FFs
obtained in the previous section at $Q^2={M_z }^2$. The
experiments showed that at this value of $Q^2$,  the interaction
between electron - positron accurse via weak interaction. In this
region the total hadron fragmentation function is given as follows
\cite{ff1,ff2}

\begin{eqnarray}
 \frac{1}{\sigma_{tot}}\frac{d\sigma^h}{dz}=F^H
(z,Q^2)=\frac{1}{\sum_ q \widehat{e}^2 _q} [2 F^H _1 (z,Q^2)+F^H
_L (z,Q^2)]\nonumber\\
\end{eqnarray}
where we have

\begin{eqnarray}
 &&2 F_1 (z,Q^2)=\sum_{q} \widehat{e}^2 [D^H_q
+D^H_{\overline{q}}](z,Q^2)\nonumber\\
&&+ \frac{\alpha_s}{2\pi} [C^1_q \otimes
(D^H_q +D^H_{\overline{q}})+C^1_g \otimes D^H_g](z,Q^2),\label{f11}
\end{eqnarray}

\begin{eqnarray}
\label{fl} F^H_L(z,Q^2)=\frac{\alpha_s}{2\pi} \sum_{q}
\widehat{e}^2 [C^L_q \otimes (D^H_q +D^H_{\overline{q}})+C^L_g
\otimes D^H_g],\nonumber\\
\end{eqnarray}

and $\widehat{e}^2 _q$ is the Electroweak  charge that is defined
as

\begin{eqnarray}
\widehat{e}^2 _q=e_q^2-2e_q \chi_1 (Q^2) V_e V_q+\chi_2(Q^2)
(1+V_e^2) (1+V_q^2),\nonumber\\
\end{eqnarray}

and the Electroweak parameters are defined as

\begin{eqnarray}
\chi_1(s)= \frac{1}{16\sin^2\theta_W\cos^2\theta_W}
       \frac{s(s-M_Z^2)}{(s-M_Z^2)^2+M_Z^2\Gamma_Z^2}, \nonumber\\
 \chi_2(s)= \frac{1}{256\sin^4\theta_W\cos^4\theta_W}
       \frac{s^2}{(s-M_Z^2)^2+M_Z^2\Gamma_Z^2}.\nonumber\\
\end{eqnarray}

\begin{eqnarray}
V_e=-1+4 \sin^2\theta_W,\nonumber\\
V_u=+1-\frac{8}{3}\sin^2\theta_W,\nonumber\\
V_d=-1+\frac{4}{3}\sin^2\theta_W.
\end{eqnarray}
The Wilson coefficients  used in Eq. (42)and Eq. (43) are defined
as follows \cite{wilson},

\begin{eqnarray}
&&C_q^1 (z)=C_F\left[(1+z^2)
\left(\frac{\ln(1-z)}{1-z}\right)_{+}-\frac{3}{2}\frac{1}{(1-z)_{+}}\right.\nonumber\\
&&\left.+2\frac{1+z^2}{1-z}\ln(z)+\frac{3}{2}(1-z)+\left(\frac{3}{2}\pi^2-\frac{9}{2}\right)
\delta(1-z) \right],\nonumber\\
\end{eqnarray}

\begin{eqnarray}
C_g^1 (z)&=&2 C_F \left[\frac{1+(1-z)^2}{z}\ln(z^2 (1-z))
-2\frac{1-z}{z}\right],
\end{eqnarray}

\begin{eqnarray}
C_q^L (z)&=&C_F,
\end{eqnarray}

\begin{eqnarray}
C_g^L (z)&=&4 C_F\frac{(1-z)}{z}.
\end{eqnarray}

The total fragmentation functions of pion, kaon and proton at the
$Q^2=M_z^2$ scale are shown in Fig. (4). We compared our result
with those from HKNS global fit ad also with data from SLD
experiment \cite{sld}. The agreement between data and our model is
quite reasonable and it means our analytical solutions are
correct.

\section{Conclusions and Remarks}
We  utilized the Laplace transform technique to calculate the
Laplace transform of splitting functions and  extract the
Fragmentation Functions of pion, kaon and proton at NLO
approximation. This technique makes this facility that the
analytical solution for the Fragmentation Functions (FFs) are
obtained more strictly by using the related kernels and we can
control the calculations in a better way.\\
 We also found a simple approach to study the
symmetry breaking in the sea quarks Fragmentation Functions. Our
results are compared with those from global fits and also with
experimental data  which indicate  good agreements between them.\\
We have also  used the HKNS code for initial input fragmentation
functions to be sure about our solutions for DGLAP evolution
equations. In a new work, we are attempting to determine  the
initial input Fragmentation Functions by Laplace transform
technique via a global fit. To do this,  we have to used available
data for total fragmentation functions and also multiplicity data.

\section*{ Acknowledgment}

This work is supported by Ferdowsi University of Mashhad under
grant 2/32653(1394/01/25). F. Taghavi Shahri thanks to professor
Firooz Arash and professor Abolfazl Mirjalili  for reading the
manuscript and for their useful comments.

\section*{Apendix A}
We present here the results for the Laplace
 transforms of   splitting  functions, denoted by $\Phi^{LO,NLO}$ and $\Theta^{LO,NLO}$ at the
NLO approximation.
\begin{widetext}

\begin{equation}
‎\Phi^{LO}_f(s)=4-\frac{8}{3}\left(\frac{1}{‎s+1‎}+‎\frac{1}{‎s+2‎}‎+2(‎\psi(s+1)+‎\gamma_E)\right)
\end{equation}‎‎
‎‎
\begin{equation}
‎\Theta‎^{LO}‎_g‎(s)=‎\frac{16}{‎3‎} ‎n_f ‎‎\left(‎‎\frac{2}{‎s‎}‎-‎\frac{2}{‎s+2‎}‎+‎\dfrac{2}{‎s+3‎}‎‎\right)‎‎‎,‎
\end{equation}‎
‎‎
\begin{equation}
‎\Theta‎^{LO}‎_f(s)=‎\frac{1}{‎s+1‎}‎-‎\frac{2}{‎s+2‎} +‎ ‎‎\frac{2}{‎s+3‎}‎,‎
\end{equation}‎
‎‎
\begin{equation}
‎ \Phi^{LO}‎_g(s)=‎\frac{33-2 n_f}{‎3‎}‎+12 ‎‎\left(‎‎\frac{1}{‎s‎}‎-‎\frac{2}{‎s+1‎}‎+‎\frac{1}{‎s+2‎}‎-‎\frac{1}{‎s+3‎}‎-‎\psi(s+1)-‎\gamma‎_E‎\right)‎,‎‎
\end{equation}

\begin{eqnarray*}
\Phi _{nsq\overline{q}}^{NLO} &=&C_{F}\left(
-\frac{C_{A}}{2}+C_{F}\right)
\left( \frac{2}{(s+1)^{3}}-\frac{2}{(s+1)^{2}}+\frac{4}{s+1}-\frac{\pi ^{2}}{%
3(s+1)}-\frac{1.9968}{(s+2)^{3}}-\frac{2}{(s+2)^{2}}\right.  \\
&&\left. +\frac{3.3246}{s+2}+\frac{3.9404}{(s+3)^{3}}-\frac{7.1312}{s+3}-%
\frac{3.602}{(s+4)^{3}}+\frac{5.8861}{s+4}+\frac{2.6484}{(s+5)^{3}}+\frac{%
3.9432}{s+5}-\frac{1.2696}{(s+6)^{3}}\right.  \\
&&\left. -\frac{14.24}{s+6}+\frac{0.2796}{(s+7)^{3}}+\frac{20.43}{s+7}-\frac{%
19.77}{s+8}+\frac{13.05}{s+9}+\frac{6.286}{s+10}+\frac{1.997}{s+11}-\frac{%
0.3076}{s+12}\right.  \\
&&\left. -2\left( \frac{4}{(s+1)^{3}}-\frac{\ln (4)}{(s+1)^{2}}-\frac{\psi (%
\frac{s}{2}+1)}{(s+1)^{2}}+\frac{\psi
(\frac{s+1}{2})}{(s+1)^{2}}+\frac{\psi
^{\prime }(\frac{s}{2}+1)}{2s+2}-\frac{\psi ^{\prime }(\frac{s+1}{2})}{2(s+1)%
}\right) \right.  \\
&&\left. -\frac{0.9984}{(s+2)^{3}}\left( \frac{16}{(s+1)^{2}}+\frac{12s}{%
(s+1)^{2}}+(s+2)\ln (16)-2(s+2)\psi (\frac{s}{2}+1)+2(s+1)\psi (\frac{s+1}{2}%
)\right. \right.  \\
&&\left. \left. +(s+2)^{2}\psi ^{\prime
}(\frac{s}{2}+1)-(s+2)^{2}\psi
^{\prime }(\frac{s+1}{2})\right) -\frac{1.9702}{(s+3)^{3}}\left( \frac{164}{%
(s+1)^{2}(s+2)^{2}}+\right. \right.  \\
&&\left. \left. \frac{284s}{(s+1)^{2}(s+2)^{2}}+\frac{188s^{2}}{%
(s+1)^{2}(s+2)^{2}}+\frac{60s^{3}}{(s+1)^{2}(s+2)^{2}}+\frac{8s^{4}}{%
(s+1)^{2}(s+2)^{2}}-\right. \right.  \\
&&\left. \left. 4(s+3)\ln (2)-2(s+3)\psi (\frac{s}{2}+1)+2(s+3)\psi (\frac{%
s+1}{2})+(s+3)^{2}\psi ^{\prime }(\frac{s}{2}+1)-\right. \right.  \\
&&\left. \left. (s+3)^{2}\psi ^{\prime }(\frac{s+1}{2})\right) -\frac{1.801}{%
(s+4)^{3}}\left( \frac{2176}{(s+1)^{2}(s+2)^{2}(s+3)^{2}}+\frac{4392s}{%
(s+1)^{2}(s+2)^{2}(s+3)^{2}}\right. \right.  \\
&&\left. \left. +\frac{3504s^{2}}{(s+1)^{2}(s+2)^{2}(s+3)^{2}}+\frac{%
1408s^{3}}{(s+1)^{2}(s+2)^{2}(s+3)^{2}}+\frac{288s^{4}}{%
(s+1)^{2}(s+2)^{2}(s+3)^{2}}\right. \right.  \\
&&\left. \left.
+\frac{24s^{5}}{(s+1)^{2}(s+2)^{2}(s+3)^{2}}+4(s+4)\ln
(2)-2(s+4)\psi (\frac{s}{2}+1)+2(s+4)\psi (\frac{s+1}{2})+\right. \right.  \\
&&\left. \left. (s+4)^{2}\psi ^{\prime
}(\frac{s}{2}+1)-(s+4)^{2}\psi
^{\prime }(\frac{s+1}{2})\right) -\frac{1.3242}{(s+5)^{3}}\left( \frac{57328%
}{(s+1)^{2}(s+2)^{2}(s+3)^{2}(s+4)^{2}}+\right. \right.  \\
&&\left. \left. \frac{146144s}{(s+1)^{2}(s+2)^{2}(s+3)^{2}(s+4)^{2}}+\frac{%
162160s^{2}}{(s+1)^{2}(s+2)^{2}(s+3)^{2}(s+4)^{2}}+\right. \right.  \\
&&\left. \left. \frac{103728s^{3}}{(s+1)^{2}(s+2)^{2}(s+3)^{2}(s+4)^{2}}+%
\frac{42144s^{4}}{(s+1)^{2}(s+2)^{2}(s+3)^{2}(s+4)^{2}}+\right. \right.  \\
&&\left. \left. \frac{11160s^{5}}{(s+1)^{2}(s+2)^{2}(s+3)^{2}(s+4)^{2}}+%
\frac{1880s^{6}}{(s+1)^{2}(s+2)^{2}(s+3)^{2}(s+4)^{2}}+\right. \right.  \\
&&\left. \left. \frac{184s^{7}}{(s+1)^{2}(s+2)^{2}(s+3)^{2}(s+4)^{2}}+\frac{%
8s^{8}}{(s+1)^{2}(s+2)^{2}(s+3)^{2}(s+4)^{2}}-4(s+5)\ln (2)\right.
\right.
\\
&&\left. \left. -2(5+s)\psi (\frac{s}{2}+1)+2(5+s)\psi (\frac{s+1}{2}%
)+(s+5)^{2}\psi ^{\prime }(\frac{s}{2}+1)-(s+5)^{2}\psi ^{\prime }(\frac{s+1%
}{2})\right) -\right.  \\
&&\left. \frac{0.6348}{(s+6)^{2}}\left( \ln (16)-2\psi
(\frac{s}{2}+4)+2\psi
(\frac{s+7}{2})+(s+6)\psi ^{\prime }(\frac{s}{2}+4)-(s+6)\psi ^{\prime }(%
\frac{s+7}{2})\right) +\right.  \\
&&\left. \frac{0.1398}{(s+7)^{2}}\left( \ln (16)+2\psi
(\frac{s}{2}+4)-2\psi
(\frac{s+9}{2})-(s+7)\psi ^{\prime }(\frac{s}{2}+4)+(s+7)\psi ^{\prime }(%
\frac{s+9}{2})\right) \right)
\end{eqnarray*}

\begin{eqnarray*}
\Phi _{nsqq}^{NLO} &=&C_{F}T_{f}\left( -\frac{2}{3(s+1)^{2}}-\frac{2}{9(s+1)}%
-\frac{2}{3(s+2)^{2}}+\frac{22}{9(s+2)}+\frac{4}{3}\psi ^{\prime
}(s+1)\right) + \\
&&C_{F}^{2}\left( \frac{5}{(s+1)^{3}}+\frac{5}{(s+1)^{2}}-\frac{5}{s+1}+%
\frac{5}{(s+2)^{3}}+\frac{3}{(s+2)^{2}}+\frac{5}{s+2}\right.  \\
&&\left. -\frac{2}{(s+1)^{2}}\left( \gamma _{E}+\frac{1}{s+1}\psi
(s+1)-(s+1)\psi ^{\prime }(s+2)\right) \right.  \\
&&\left. -\frac{2}{(s+2)^{2}}\left( \gamma _{E}+\frac{1}{s+2}\psi
(s+2)-(s+2)\psi ^{\prime }(s+3)\right) \right.  \\
&&\left. +4\left( \left( \psi (s+1)+\gamma _{E}\right) \psi ^{\prime }(s+1)-%
\frac{1}{2}\psi ^{\prime \prime }(s+1)\right) -3\psi ^{\prime
}(s+1)+4\psi
^{\prime \prime }(s+1)\right)  \\
&&+C_{A}C_{F}\left( -\frac{1}{(s+1)^{3}}+\frac{5}{6(s+1)^{2}}+\frac{53}{%
18(s+1)}+\frac{\pi ^{2}}{6(s+1)}-\frac{1}{(s+2)^{3}}\right.  \\
&&\left. +\frac{5}{6(s+2)^{2}}-\frac{187}{18(s+2)}+\frac{\pi ^{2}}{6(s+2)}-%
\frac{67}{9}\left( \psi (s+1)+\gamma _{E}\right) +\frac{1}{3}\pi
^{2}\right.
\\
&&\left. \left( \psi (s+1)+\gamma _{E}\right) +2\left( \frac{67}{18}-\frac{%
\pi ^{2}}{6}\right) \left( \psi (s+1)+\gamma _{E}\right)
-\frac{11}{3}\psi ^{\prime }(s+1)-\psi ^{\prime \prime
}(s+1)\right)
\end{eqnarray*}

\begin{eqnarray*}
\Phi _{q}^{NLO} &=&C_{F}T_{f}\left( -\frac{40}{9s}+\frac{4}{(s+1)^{3}}+\frac{%
28}{3(s+1)^{2}}-\frac{146}{9(s+1)}+\frac{4}{(s+2)^{3}}+\frac{52}{3(s+2)^{2}}+%
\frac{94}{9(s+2)}+\right.  \\
&&\left. \frac{16}{3(s+3)^{2}}+\frac{112}{9(s+3)}+\frac{4}{3}\psi
^{\prime
}(s+1)\right) +C_{F}^2\left( \frac{7}{(s+1)^{3}}+\frac{3}{(s+1)^{2}}-\frac{1}{%
s+1}-\frac{\pi ^{2}}{3(s+1)}+\right.  \\
&&\left. \frac{3.0032}{(s+2)^{3}}+\frac{1}{(s+2)^{2}}+\frac{8.3246}{s+2}+%
\frac{3.9404}{(s+3)^{3}}-\frac{7.1312}{s+3}-\frac{3.602}{(s+4)^{3}}+\frac{%
5.886}{s+4}+\frac{2.6484}{(s+5)^{3}}\right.  \\
&&\left. +\frac{3.9432}{s+5}-\frac{1.2696}{(s+6)^{3}}-\frac{14.2478}{s+6}+%
\frac{0.2796}{(s+7)^{3}}+\frac{20.4376}{s+7}-\frac{19.7727}{s+8}+\frac{13.056%
}{s+9}-\frac{6.2862}{s+10}\right.  \\
&&\left. +\frac{1.9971}{s+11}-\frac{0.3075}{s+12}-\frac{8}{(s+1)^{3}}+\frac{%
2\ln (4)}{(s+1)^{2}}+\frac{2\psi (\frac{s}{2}+1)}{(s+1)^{2}}-\frac{2\psi (%
\frac{s+1}{2})}{(s+1)^{2}}-\frac{\psi ^{\prime }(\frac{s}{2}+1)}{s+1}%
+\right.  \\
&&\left. \frac{\psi ^{\prime }(\frac{s+1}{2})}{(s+1)^{2}}-\frac{0.9984}{%
(s+2)^{3}}\left(
\frac{16}{(s+1)^{2}}+\frac{12s}{(s+1)^{2}}+(s+2)\ln
(16)-2(s+2)\psi (\frac{s}{2}+1)+\right. \right.  \\
&&\left. \left. 2(s+2)\psi (\frac{s+1}{2})+(s+2)^{2}\psi ^{\prime }(\frac{s}{%
2}+1)-(s+2)^{2}\psi ^{\prime }(\frac{s+1}{2})\right)- \frac{1.9702}{(s+3)^{3}}%
\left( \frac{164}{(s+1)^{2}(s+2)^{2}}\right. \right.  \\
&&\left. \left. +\frac{284s}{(s+1)^{2}(s+2)^{2}}+\frac{188s^{2}}{%
(s+1)^{2}(s+2)^{2}}+\frac{60s^{3}}{(s+1)^{2}(s+2)^{2}}+\frac{8s^{4}}{%
(s+1)^{2}(s+2)^{2}}-\right. \right.  \\
&&\left. \left. 4(s+3)\ln (2)-2(s+3)\psi (\frac{s}{2}+1)+2(s+3)\psi (\frac{%
s+1}{2})+(s+3)^{2}\psi ^{\prime }(\frac{s}{2}+1)-(s+3)^{2}\psi ^{\prime }(%
\frac{s+1}{2})\right) \right.  \\
&&\left. -\frac{1.801}{(s+4)^{3}}\left( \frac{2176}{%
(s+1)^{2}(s+2)^{2}(s+3)^{2}}+\frac{4392s}{(s+1)^{2}(s+2)^{2}(s+3)^{2}}+\frac{%
3504s^{2}}{(s+1)^{2}(s+2)^{2}(s+3)^{2}}+\right. \right.  \\
&&\left. \left. \frac{1408s^{3}}{(s+1)^{2}(s+2)^{2}(s+3)^{2}}+\frac{288s^{4}%
}{(s+1)^{2}(s+2)^{2}(s+3)^{2}}+\frac{24s^{5}}{(s+1)^{2}(s+2)^{2}(s+3)^{2}}%
+\right. \right.  \\
&&\left. \left. 4(s+4)\ln (2)-2(s+4)\psi (\frac{s}{2}+1)+2(s+4)\psi (\frac{%
s+1}{2})+(s+4)^{2}\psi ^{\prime }(\frac{s}{2}+1)-(s+4)^{2}\psi ^{\prime }(%
\frac{s+1}{2})\right) \right.  \\
&&\left.- \frac{1.3242}{(s+5)^{3}}\left( \frac{57328}{%
(s+1)^{2}(s+2)^{2}(s+3)^{2}(s+4)^{2}}+\frac{146144s}{%
(s+1)^{2}(s+2)^{2}(s+3)^{2}(s+4)^{2}}+\right. \right.  \\
&&\left. \left. \frac{162160s^{2}}{(s+1)^{2}(s+2)^{2}(s+3)^{2}(s+4)^{2}}+%
\frac{103728s^{3}}{(s+1)^{2}(s+2)^{2}(s+3)^{2}(s+4)^{2}}+\right. \right.  \\
&&\left. \left. \frac{42144s^{4}}{(s+1)^{2}(s+2)^{2}(s+3)^{2}(s+4)^{2}}+%
\frac{11160s^{5}}{(s+1)^{2}(s+2)^{2}(s+3)^{2}(s+4)^{2}}+\right. \right.  \\
&&\left. \left. \frac{1880s^{6}}{(s+1)^{2}(s+2)^{2}(s+3)^{2}(s+4)^{2}}+\frac{%
184s^{7}}{(s+1)^{2}(s+2)^{2}(s+3)^{2}(s+4)^{2}}+\right. \right.  \\
&&\left. \left. \frac{8s^{8}}{(s+1)^{2}(s+2)^{2}(s+3)^{2}(s+4)^{2}}%
-4(s+5)\ln (2)-2(s+5)\psi (\frac{s}{2}+1)+2(s+5)\psi
(\frac{s+1}{2})\right.
\right.  \\
&&\left. \left. +(s+5)^{2}\psi ^{\prime
}(\frac{s}{2}+1)-(s+5)^{2}\psi
^{\prime }(\frac{s+1}{2})\right) -\frac{2}{(s+1)^{2}}(\gamma _{E}+\frac{1}{%
s+1}+\psi (s+1)-(s+1)\psi ^{\prime }(s+2))\right.  \\
&&\left. -\frac{2}{(s+2)^{2}}(\gamma _{E}+\frac{1}{s+2}+\psi
(s+2)-(s+2)\psi
^{\prime }(s+3))-\frac{0.6348}{(s+6)^{2}}\left( \ln (16)-2\psi (\frac{s}{2}%
+4)+\right. \right.  \\
&&\left. \left. 2\psi (\frac{s+7}{2})+(s+6)\psi ^{\prime }(\frac{s}{2}%
+4)-(s+6)\psi ^{\prime }(\frac{s+7}{2})\right) +\frac{0.1398}{(s+7)^{2}}%
\left( \ln (16)+2\psi (\frac{s}{2}+4)-\right. \right.  \\
&&\left. \left. 2\psi (\frac{s+9}{2})-(s+7)\psi ^{\prime }(\frac{s}{2}%
+4)+(s+7)\psi ^{\prime }(\frac{s+9}{2})\right) +\right.  \\
&&\left. 4\left( (\psi (s+1)+\gamma _{E})\psi ^{\prime }(s+1)-\frac{1}{2}%
\psi ^{\prime \prime }(s+1)\right) -3\psi ^{\prime }(s+1)+4\psi
^{\prime
\prime }(s+1)\right) + \\
\end{eqnarray*}
\begin{eqnarray*}
&&C_{A}C_{F}\left( \frac{2}{(s+1)^{3}}+\frac{11}{6(s+1)^{2}}+\frac{17}{%
18(s+1)}+\frac{\pi ^{2}}{3(s+1)}-\frac{0.0016}{(s+2)^{3}}+\frac{11}{%
6(s+2)^{2}}-\frac{10.4062}{s+2}\right.  \\
&&\left. -\frac{1.9702}{(s+3)^{3}}+\frac{3.5656}{s+3}+\frac{1.801}{(s+4)^{3}}%
-\frac{2.9430}{s+4}-\frac{1.3242}{(s+5)^{3}}-\frac{1.9716}{s+5}+\frac{0.6348%
}{(s+6)^{3}}+\frac{7.1239}{s+6}\right.  \\
&&\left. -\frac{0.1398}{(s+7)^{3}}-\frac{10.2188}{s+7}+\frac{9.8863}{s+8}-%
\frac{6.5284}{s+9}+\frac{3.1431}{s+10}-\frac{0.9985}{s+11}+\frac{0.1537}{s+12%
}-\right.  \\
&&\left. \frac{67(\psi (s+1)+\gamma _{E})}{9}+\frac{1}{3}\pi
^{2}(\psi (s+1)+\gamma _{E})+2\left( \frac{67}{18}-\frac{\pi
^{2}}{6}\right) (\psi
(s+1)+\gamma _{E})-\frac{\ln (4)}{(s+1)^{2}}-\right.  \\
&&\left. \frac{\psi (\frac{s}{2}+1)}{(s+1)^{2}}+\frac{\psi (\frac{s+1}{2})}{%
(s+1)^{2}}+\frac{\psi ^{\prime }(\frac{s}{2}+1)}{2s+2}-\frac{\psi ^{\prime }(%
\frac{s+1}{2})}{2s+2}+\frac{0.4992}{(s+2)^{3}}\left( \frac{16}{(s+1)^{2}}+%
\frac{12s}{(s+1)^{2}}+(s+2)\ln (16)\right. \right.  \\
&&\left. \left. -2(s+2)\psi (\frac{s}{2}+1)+2(s+2)\psi (\frac{s+1}{2}%
)+(s+2)^{2}\psi ^{\prime }(\frac{s}{2}+1)-(s+2)^{2}\psi ^{\prime }(\frac{s+1%
}{2})\right) +\right.  \\
&&\left. \frac{0.9851}{(s+3)^{3}}\left( \frac{164}{(s+1)^{2}(s+2)^{2}}+\frac{%
284s}{(s+1)^{2}(s+2)^{2}}+\frac{188s^{2}}{(s+1)^{2}(s+2)^{2}}+\frac{60s^{3}}{%
(s+1)^{2}(s+2)^{2}}+\right. \right.  \\
&&\left. \left. \frac{8s^{4}}{(s+1)^{2}(s+2)^{2}}-4(s+3)\ln (2)-2(s+3)\psi (%
\frac{s}{2}+1)+2(s+3)\psi (\frac{s+1}{2})+\right. \right.  \\
&&\left. \left. (s+3)^{2}\psi ^{\prime
}(\frac{s}{2}+1)-(s+3)^{2}\psi
^{\prime }(\frac{s+1}{2})\right) +\frac{0.9005}{(s+4)^{3}}\right.  \\
&&\left. \left( \frac{2176}{(s+1)^{2}(s+2)^{2}(s+3)^{2}}+\frac{4392s}{%
(s+1)^{2}(s+2)^{2}(s+3)^{2}}+\frac{3504s^{2}}{(s+1)^{2}(s+2)^{2}(s+3)^{2}}%
+\right. \right.  \\
&&\left. \left. \frac{1408s^{3}}{(s+1)^{2}(s+2)^{2}(s+3)^{2}}+\frac{288s^{4}%
}{(s+1)^{2}(s+2)^{2}(s+3)^{2}}+\frac{24s^{5}}{(s+1)^{2}(s+2)^{2}(s+3)^{2}}%
+\right. \right.  \\
&&\left. \left. 4(s+4)\ln (2)-2(s+4)\psi (\frac{s}{2}+1)+2(s+4)\psi (\frac{%
s+1}{2})+(s+4)^{2}\psi ^{\prime }(\frac{s}{2}+1)-(s+4)^{2}\psi ^{\prime }(%
\frac{s+1}{2})\right) \right.  \\
&&\left. +\frac{0.6621}{(s+5)^{3}}\left( \frac{57328}{%
(s+1)^{2}(s+2)^{2}(s+3)^{2}(s+4)^{2}}+\frac{146144s}{%
(s+1)^{2}(s+2)^{2}(s+3)^{2}(s+4)^{2}}+\right. \right.  \\
&&\left. \left. \frac{162160s^{2}}{(s+1)^{2}(s+2)^{2}(s+3)^{2}(s+4)^{2}}+%
\frac{103728s^{3}}{(s+1)^{2}(s+2)^{2}(s+3)^{2}(s+4)^{2}}+\right. \right.  \\
&&\left. \left. \frac{42144s^{4}}{(s+1)^{2}(s+2)^{2}(s+3)^{2}(s+4)^{2}}+%
\frac{11160s^{5}}{(s+1)^{2}(s+2)^{2}(s+3)^{2}(s+4)^{2}}+\right. \right.  \\
&&\left. \left. \frac{1880s^{6}}{(s+1)^{2}(s+2)^{2}(s+3)^{2}(s+4)^{2}}+\frac{%
184s^{7}}{(s+1)^{2}(s+2)^{2}(s+3)^{2}(s+4)^{2}}+\right. \right.  \\
&&\left. \left. \frac{8s^{8}}{(s+1)^{2}(s+2)^{2}(s+3)^{2}(s+4)^{2}}%
-4(s+5)\ln (2)-2(s+5)\psi (\frac{s}{2}+1)+2(s+5)\psi
(\frac{s+1}{2})+\right.
\right.  \\
&&\left. \left. (s+5)^{2}\psi ^{\prime
}(\frac{s}{2}+1)-(s+5)^{2}\psi ^{\prime }(\frac{s+1}{2})\right)
+\frac{0.3174}{(s+6)^{2}}\left( \ln
(16)-2\psi (\frac{s}{2}+4)+2\psi (\frac{s+7}{2})+\right. \right.  \\
&&\left. \left. (s+6)\psi ^{\prime }(\frac{s}{2}+4)-(s+6)\psi ^{\prime }(%
\frac{s+7}{2})\right) -\frac{0.0699}{(s+7)^{2}}\left( \ln (16)+2\psi (\frac{s%
}{2}+4)-2\psi (\frac{s+9}{2})-\right. \right.  \\
&&\left. \left. (s+7)\psi ^{\prime }(\frac{s}{2}+4)+(s+7)\psi ^{\prime }(%
\frac{s+9}{2})\right) -\frac{11}{3}\psi ^{\prime }(s+1)-\psi
^{\prime \prime }(s+1)\right)
\end{eqnarray*}

\begin{eqnarray*}
\Theta _{q}^{NLO} &=&T_{f}^2\left( \frac{8}{3(s+1)^{2}}-\frac{40}{9(s+1)}-%
\frac{16}{3(s+2)^{2}}+\frac{32}{9(s+2)}+\frac{16}{3(s+3)^{ý2ý}}-\frac{32}{%
9(s+3)}+\right.  \\
&&\left. \frac{8(\psi (s+2)+\gamma _{E})}{3(s+1)}-\frac{16(\psi
(s+3)+\gamma
_{E})}{3(s+2)}+\frac{16(\psi (s+4)+\gamma _{E})}{3(s+3)}\right) + \\
&&C_{A}T_{f}\left( -\frac{40}{9s}+\frac{4}{(s+1)^{3}}+\frac{8}{3(s+1)^{2}}+%
\frac{26}{9(s+1)}+\frac{24}{(s+2)^{3}}+\frac{68}{3(s+2)^{2}}-\frac{33.231}{%
s+2}\right.  \\
&&\left. -\frac{4\pi ^{2}}{3(s+2)}+\frac{8}{3(s+3)^{2}}+\frac{96.875}{s+3}-%
\frac{67.644}{s+4}+\frac{83.04}{s+5}-\frac{82.976}{s+6}+\frac{56.16}{s+7}-%
\frac{22}{s+8}\right.  \\
&&\left. -\frac{22(\psi (s+2)+\gamma _{E})}{3(s+1)}+\frac{20(\psi
(s+3)+\gamma _{E})}{3(s+2)}-\frac{20(\psi (s+4)+\gamma
_{E})}{3(s+3)}\right.
\\
&&\left. -2\left( \frac{4}{(s+1)^{3}}-\frac{\ln (4)}{(s+1)^{2}}-\frac{\psi (%
\frac{s}{2}+1)}{(s+1)^{2}}+\frac{\psi
(\frac{s+1}{2})}{(s+1)^{2}}+\frac{\psi
^{\prime }(\frac{s}{2}+1)}{2s+2}-\frac{\psi ^{\prime }(\frac{s+1}{2})}{2(s+1)%
}\right) +\right.  \\
&&\left. \frac{2}{(s+2)^{3}}\left( \frac{16}{(s+1)^{2}}+\frac{12s}{(s+1)^{2}}%
+(s+2)\ln (16)-2(s+2)\psi (\frac{s}{2}+1)+2(s+2)\psi
(\frac{s+1}{2})\right.
\right.  \\
&&\left. \left. +(s+2)^{2}\psi ^{\prime
}(\frac{s}{2}+1)-(s+2)^{2}\psi
^{\prime }(\frac{s+1}{2})\right) -\frac{2}{(s+3)^{3}}\left( \frac{164}{%
(s+1)^{2}(s+2)^{2}}+\frac{284s}{(s+1)^{2}(s+2)^{2}}\right. \right.  \\
&&\left. \left. +\frac{188s^{2}}{(s+1)^{2}(s+2)^{2}}+\frac{60s^{3}}{%
(s+1)^{2}(s+2)^{2}}+\frac{8s^{4}}{(s+1)^{2}(s+2)^{2}}-4(s+3)\ln
(2)-2(s+3)\psi (\frac{s}{2}+1)ý+ý\right. \right.  \\
&&\left. \left. 2(s+3)\psi (\frac{s+1}{2})+(s+ý3ý)^{2}\psi ^{\prime }(\frac{s}{%
2}+1)-(s+3)^{2}\psi ^{\prime }(\frac{s+1}{2})\right) +\frac{2(\pi
^{2}+6(\psi (s+2)+\gamma _{E})^{2}-6\psi ^{\prime }(s+2))}{6s+6}\right.  \\
&&\left. -\frac{8}{(s+1)^{2}}\left( \gamma _{E}+\frac{1}{s+1}+\psi
(s+1)-(s+1)\psi ^{\prime }(s+2)\right) -\frac{4(\pi ^{2}+6(\psi
(s+3)+\gamma
_{E})^{2}-6\psi ^{\prime }(s+3))}{6s+12}\right.  \\
&&\left. +\frac{16}{(s+2)^{2}}\left( \gamma
_{E}+\frac{1}{s+2}+\psi (s+2)-(s+2)\psi ^{\prime }(s+3)\right)
+\frac{4(\pi ^{2}+6(\psi (s+4)+\gamma
_{E})^{2}-6\psi ^{\prime }(s+4))}{6s+18}\right.  \\
&&\left. -\frac{16}{(s+3)^{2}}\left( \gamma
_{E}+\frac{1}{s+3}+\psi
(s+3)-(s+3)\psi ^{\prime }(s+4)\right) \right) + \\
&&C_{F}T_{f}\left( -\frac{2}{(s+1)^{3}}+\frac{7}{(s+1)^{2}}-\frac{12}{s+1}-%
\frac{2\pi ^{2}}{3(s+1)}+\frac{4}{(s+2)^{3}}-\frac{8}{(s+2)^{2}}+\frac{39.16%
}{s+2}+\frac{4\pi ^{2}}{3(s+2)}-\frac{8}{(s+ý3ý)^{3}}\right.  \\
&&\left. -\frac{65.856}{s+3}-\frac{4\pi ^{2}}{3(s+3)}+\frac{77.872}{s+4}-%
\frac{81.216}{s+5}+\frac{80.128}{s+6}-\frac{51.968}{s+7}+\frac{17.6}{s+8}+%
\frac{2(\psi (s+1)+\gamma _{E})}{s+1}+\right.  \\
&&\left. \frac{4(\psi (s+2)+\gamma _{E})}{s+1}-\frac{4(\psi
(s+2)+\gamma _{E})}{s+2}+\frac{4(\psi (s+3)+\gamma
_{E})}{s+3}-\frac{2(\pi ^{2}+6(\psi
(s+2)+\gamma _{E})^{2}-6\psi ^{\prime }(s+2))}{6s+6}\right.  \\
&&\left. +\frac{12}{(s+1)^{2}}\left( \gamma
_{E}+\frac{1}{s+1}+\psi (s+1)-(s+1)\psi ^{\prime }(s+2)\right)
+\frac{4(\pi ^{2}+6(\psi (s+3)+\gamma
_{E})^{2}-6\psi ^{\prime }(s+3))}{6s+12}\right.  \\
&&\left. -\frac{24}{(s+2)^{2}}\left( \gamma
_{E}+\frac{1}{s+2}+\psi (s+2)-(s+2)\psi ^{\prime }(s+3)\right)
-\frac{4(\pi ^{2}+6(\psi (s+4)+\gamma
_{E})^{2}-6\psi ^{\prime }(s+4))}{6s+18}+\right.  \\
&&\left. \frac{24}{(s+3)^{2}}\left( \gamma _{E}+\frac{1}{s+3}+\psi
(s+3)-(s+3)\psi ^{\prime }(s+4)\right) \right)
\end{eqnarray*}

\begin{eqnarray*}
\Theta _{g}^{NLO} &=&C_{F}^{2}\left( \frac{2}{(s+1)^{3}}+\frac{8}{(s+1)^{2}}-%
\frac{16.66}{s+1}-\frac{1}{(s+2)^{3}}-\frac{1}{2(s+2)^{2}}+\frac{34.196}{s+2}%
-\frac{40.096}{s+3}+\right.  \\
&&\left. \frac{42.432}{s+4}-\frac{35.224}{s+5}+\frac{17.392}{s+6}-\frac{4.4}{%
s+7}-\frac{2(\psi (s+3)+\gamma _{E})}{s+2}+\right.  \\
&&\left. \frac{1}{3s}\left( \pi ^{2}+6(\psi (s+1)+\gamma
_{E})^{2}-6\psi ^{\prime }(s+1)\right) -\frac{8}{s^{3}}(1+s\gamma
_{E}+s(\psi (s)-s\psi
^{\prime }(s+1)))-\right.  \\
&&\left. \frac{2}{6s+6}\left( \pi ^{2}+6\left( \psi (s+2)+\gamma
_{E}\right)
^{2}-6\psi ^{\prime }(s+2)\right) +\frac{8}{(s+1)^{2}}\left( \gamma _{E}+%
\frac{1}{s+1}+\psi (s+1)-\right. \right.  \\
&&\left. \left. (s+1)\psi ^{\prime }(s+2)\right)
+\frac{1}{6s+12}\left( \pi
^{2}+6(\psi (s+3)+\gamma _{E})^{2}-6\psi ^{\prime }(s+3)\right) -\right.  \\
&&\left. \frac{4}{(s+2)^{2}}\left( \gamma _{E}+\frac{1}{s+2}+\psi
(s+2)-(s+2)\psi ^{\prime }(s+3)\right) \right) + \\
&&C_{A}C_{F}\left(
-\frac{4}{s^{3}}+\frac{6}{s^{2}}+\frac{17}{9s}-\frac{2\pi
^{2}}{3s}-\frac{8}{(s+1)^{2}}+\frac{25.2}{s+1}-\frac{4}{(s+2)^{3}}-\frac{9}{%
(s+2)^{2}}-\frac{23.27}{s+2}-\right.  \\
&&\left. \frac{\pi ^{2}}{3(s+2)}-\frac{8}{3(s+3)^{2}}+\frac{35.99}{s+3}-%
\frac{41.046}{s+4}+\frac{35.01}{s+5}-\frac{17.444}{s+6}+\frac{3.3}{s+7}+%
\frac{2(\psi (s+3)+\gamma _{E})}{s+2}\right.  \\
&&\left. +\frac{1}{s^{2}}\left( \ln (16)-2\psi (\frac{s}{2}+1)+2\psi (\frac{%
s+1}{2})+s\psi ^{\prime }(\frac{s}{2}+1)-s\psi ^{\prime }(\frac{s+1}{2}%
)\right) -\right.  \\
&&\left. 2\left( \frac{4}{(s+1)^{3}}-\frac{\ln (4)}{(s+1)^{2}}-\frac{\psi (%
\frac{s}{2}+1)}{(s+1)^{2}}+\frac{\psi
(\frac{s+1}{2})}{(s+1)^{2}}+\frac{\psi
^{\prime }(\frac{s}{2}+1)}{2s+2}-\frac{\psi ^{\prime }(\frac{s+1}{2})}{2s+2}%
\right) +\right.  \\
&&\left. \frac{1}{2(s+2)^{3}}\left( \frac{16}{(s+1)^{2}}+\frac{12s}{(s+1)^{2}%
}+(s+2)\ln (16)-2(s+2)\psi (\frac{s}{2}+1)+2(s+2)\psi (\frac{s+1}{2}%
)+\right. \right.  \\
&&\left. \left. (s+2)^{2}\psi ^{\prime
}(\frac{s}{2}+1)-(s+2)^{2}\psi
^{\prime }(\frac{s+1}{2})\right) -\frac{1}{3s}\right.  \\
&&\left. \left( \pi ^{2}+6\left( \psi (s+1)+\gamma _{E}\right)
^{2}-6\psi ^{\prime }(s+1)\right) +\frac{12}{s^{3}}(1+s\gamma
_{E}+s(\psi (s)-s\psi
^{\prime }(s+1)))+\right.  \\
&&\left. \frac{2}{6s+6}\left( \pi ^{2}+6\left( \psi (s+2)+\gamma
_{E}\right)
^{2}-6\psi ^{\prime }(s+2)\right) -\right.  \\
&&\left. \frac{12}{(s+1)^{2}}\left( \gamma _{E}+\frac{1}{s+1}+\psi
(s+1)-(s+1)\psi ^{\prime }(s+2)\right) -\right.  \\
&&\left. \frac{1}{6s+12}\left( \pi ^{2}+6\left( \psi (s+3)+\gamma
_{E}\right) ^{2}-6\psi ^{\prime }(s+3)\right) +\right.  \\
&&\left. \frac{6}{(s+2)^{2}}\left( \gamma _{E}+\frac{1}{s+2}+\psi
(s+2)-(s+2)\psi ^{\prime }(s+3)\right) \right)
\end{eqnarray*}

\begin{eqnarray*}
\Phi _{g}^{NLO} &=&C_{F}T_{f}\left( -\frac{16}{3s^{2}}+\frac{92}{9s}+\frac{4%
}{(s+1)^{3}}-\frac{10}{(s+1)^{2}}-\frac{4}{s+1}+\frac{4}{(s+2)^{3}}-\frac{14%
}{(s+2)^{2}}+\frac{12}{s+2}-\frac{16}{3(s+3)^{2}}-\frac{164}{9(s+3)}\right)  \\
&&+C_{A}T_{f}\left( \frac{8}{3s^{2}}-\frac{46}{9s}-\frac{4}{(s+1)^{2}}+\frac{%
58}{9(s+1)}+\frac{4}{(s+2)^{2}}-\frac{38}{9(s+2)}-\frac{8}{3(s+3)^{2}}+\frac{%
46}{9(s+3)}\right.  \\
&&\left. +\frac{8}{3}\psi ^{\prime }(s+1)\right) +C_{A}^{2}\left( -\frac{8}{%
s^{3}}+\frac{22}{3s^{2}}+\frac{2}{(s+1)^{3}}+\frac{11}{(s+1)^{2}}+\frac{%
4.4407}{s+1}-\frac{17.9984}{(s+2)^{3}}+\frac{1}{(s+2)^{2}}-\right.  \\
&&\left. \frac{6.9024}{s+2}-\frac{\pi ^{2}}{3(s+2)}+\frac{5.9702}{(s+3)^{3}}+%
\frac{22}{3(s+3)^{2}}-\frac{6.7917}{s+3}+\frac{\pi ^{2}}{3(s+3)}-\frac{1.801%
}{(s+4)^{3}}-\frac{3.5389}{s+4}+\right.  \\
&&\left. \frac{1.3242}{(s+5)^{3}}+\frac{1.2736}{s+5}-\frac{0.6348}{(s+6)^{3}}%
-\frac{5.6479}{s+6}+\frac{0.1398}{(s+7)^{3}}+\frac{9.2228}{s+7}-\frac{7.6863%
}{s+8}+\frac{6.5284}{s+9}-\frac{3.1431}{s+10}\right.  \\
&&\left. +\frac{0.9985}{s+11}-\frac{0.1537}{s+12}-\frac{67(\psi
(s+1)+\gamma _{E})}{9}+\frac{1}{3}\pi^2(\psi (s+1)+\gamma
_{E})-\frac{1}{s^{2}}\left( \ln
(16)-2\psi (\frac{s}{2}+1)+\right. \right.  \\
&&\left. \left. 2\psi (\frac{s+1}{2})+s\psi ^{\prime
}(\frac{s}{2}+1)-s\psi
^{\prime }(\frac{s+1}{2})\right) +2\left( \frac{4}{(s+1)^{3}}-\frac{\ln (4)}{%
(s+1)^{2}}-\frac{\psi (\frac{s}{2}+1)}{(s+1)^{2}}+\frac{\psi (\frac{s+1}{2})%
}{(s+1)^{2}}+\right. \right.  \\
&&\left. \left. \frac{\psi ^{\prime
}(\frac{s}{2}+1)}{2s+2}-\frac{\psi ^{\prime
}(\frac{s+1}{2})}{2s+2}\right) -\frac{1.9992}{(s+2)^{3}}\left(
\frac{16}{(s+1)^{2}}+\frac{12s}{(s+1)^{2}}+(s+2)\ln (16)-2(s+2)\psi (\frac{s%
}{2}+1)\right. \right.  \\
&&\left. \left. +2(s+2)\psi (\frac{s+1}{2})+(s+2)^{2}\psi ^{\prime }(\frac{s%
}{2}+1)-(s+2)^{2}\psi ^{\prime }(\frac{s+1}{2})\right) +\frac{0.0149}{%
(s+1)^{3}}\left( \frac{164}{(s+1)^{2}(s+2)^{2}}\right. \right.  \\
&&\left. \left. +\frac{284s}{(s+1)^{2}(s+2)^{2}}+\frac{188s^{2}}{%
(s+1)^{2}(s+2)^{2}}+\frac{60s^{3}}{(s+1)^{2}(s+2)^{2}}+\frac{8s^{4}}{%
(s+1)^{2}(s+2)^{2}}-4(s+3)\ln (2)-\right. \right.  \\
&&\left. \left. 2(s+3)\psi (\frac{s}{2}+1)+2(s+3)\psi (\frac{s+1}{2}%
)+(s+3)^{2}\psi ^{\prime }(\frac{s}{2}+1)-(s+3)^{2}\psi ^{\prime }(\frac{s+1%
}{2})\right) -\frac{0.9005}{(s+4)^{3}}\right.  \\
&&\left. \left( \frac{2176}{(s+1)^{2}(s+2)^{2}(s+3)^{2}}+\frac{4392s}{%
(s+1)^{2}(s+2)^{2}(s+3)^{2}}+\frac{3504s^{2}}{(s+1)^{2}(s+2)^{2}(s+3)^{2}}%
\right. \right.  \\
&&\left. \left. +\frac{1408s^{3}}{(s+1)^{2}(s+2)^{2}(s+3)^{2}}+\frac{288s^{4}%
}{(s+1)^{2}(s+2)^{2}(s+3)^{2}}+\frac{24s^{5}}{(s+1)^{2}(s+2)^{2}(s+3)^{2}}%
+\right. \right.  \\
&&\left. \left. 4(s+4)\ln (2)-2(s+4)\psi (\frac{s}{2}+1)+2(s+4)\psi (\frac{%
s+1}{2})+(s+4)^{2}\psi ^{\prime }(\frac{s}{2}+1)-(s+4)^{2}\psi ^{\prime }(%
\frac{s+1}{2})\right) \right.  \\
&&\left. -\frac{0.6621}{(s+5)^{3}}\left( \frac{57328}{%
(s+1)^{2}(s+2)^{2}(s+3)^{2}(s+4)^{2}}+\frac{146144s}{%
(s+1)^{2}(s+2)^{2}(s+3)^{2}(s+4)^{2}}\right. \right.  \\
&&\left. +\frac{162160s^{2}}{(s+1)^{2}(s+2)^{2}(s+3)^{2}(s+4)^{2}}\left. +%
\frac{103728s^{3}}{(s+1)^{2}(s+2)^{2}(s+3)^{2}(s+4)^{2}}\right. \right.  \\
&&\left. \left. +\frac{42144s^{4}}{(s+1)^{2}(s+2)^{2}(s+3)^{2}(s+4)^{2}}\left. +%
\frac{11160s^{5}}{(s+1)^{2}(s+2)^{2}(s+3)^{2}9s+4)^{2}}\right.
\right.
\right.  \\
&&\left. \left. +\frac{1880s^{6}}{(s+1)^{2}(s+2)^{2}(s+3)^{2}(s+4)^{2}}+\frac{%
184s^{7}}{(s+1)^{2}(s+2)^{2}(s+3)^{2}(s+4)^{2}}+\frac{8s^{8}}{%
(s+1)^{2}(s+2)^{2}(s+3)^{2}(s+4)^{2}}\right. \right.  \\
&&\left. -4(s+5)\ln (2)-2(s+5)\psi (\frac{s}{2}+1)+2(s+5)\psi (\frac{s+1}{2}%
)+(s+5)^{2}\psi ^{\prime }(\frac{s}{2}+1)-(s+5)^{2}\psi ^{\prime }(\frac{s+1%
}{2})\right)  \\
&&\left. +\frac{4}{s^{3}}(1+\gamma _{E}s+s(\psi (s)-s\psi ^{\prime }(s+1)))-%
\frac{8}{(s+1)^{2}}(\gamma _{E}+\frac{1}{s+1}+\psi (s+1)-(s+1)\psi
^{\prime
}(s+2))+\right.  \\
&&\left. \frac{4}{(s+2)^{2}}(\gamma _{E}+\frac{1}{s+2}+\psi
(s+2)-(s+2)\psi ^{\prime }(s+3))-\frac{4}{(s+3)^{2}}(\gamma
_{E}+\frac{1}{s+3}+\psi
(s+3)-(s+3)\psi ^{\prime }(s+4))\right)  \\
&&\left. -\frac{0.3174}{(s+6)^{2}}\left( \ln (16)-2\psi
(\frac{s}{2}+4)+2\psi
(\frac{s+7}{2})+(s+6)\psi ^{\prime }(\frac{s}{2}+4)-(s+6)\psi ^{\prime }(%
\frac{s+7}{2})\right) +\right.  \\
&&\left. \frac{0.0699}{(s+7)^{2}}\left( \ln (16)+2\psi
(\frac{s}{2}+4)-2\psi
(\frac{s+9}{2})-(s+7)\psi ^{\prime }(\frac{s}{2}+4)+(s+7)\psi ^{\prime }(%
\frac{s+9}{2})\right) \right.  \\
&&\left. +4\left( (\psi (s+1)+\gamma _{E})\psi ^{\prime }(s+1)-\frac{1}{2}%
\psi ^{\prime \prime }(s+1)\right) -\frac{22}{3}\psi ^{\prime
}(s+1)+3\psi ^{\prime \prime }(s+1)\right)
\end{eqnarray*}

\begin{eqnarray*}
C_{q}^{L} &=&\frac{C_{F}}{s+1} \\
C_{g}^{L} &=&4C_{F}\left( \frac{1}{s}-\frac{1}{s+1}\right)  \\
C_{q}^{1} &=&\frac{-9}{2}+\frac{2\pi ^{2}}{3}+\frac{2}{3(s+1)}-\frac{2}{%
3(s+2)}-2\psi ^{\prime }(s+1)-2\psi ^{\prime }(s+3) \\
C_{g}^{1} &=&2C_{F}\left(
-\frac{2}{s}+\frac{2}{s+1}-\frac{2(2+s(\psi
(s+1)+\gamma _{E}))}{s^{2}}+\frac{2(2+(1+s)(\psi (s+1)+\gamma _{E}))}{%
(s+1)^{2}}\right.  \\
&&\left. \frac{2+(s+2)(\psi (s+3)+\gamma _{E})}{(s+2)^{2}}\right)
\end{eqnarray*}

\end{widetext}
\section*{Appendix B}
We bring in below the coefficients of singlet and gluon distributions of Eq. (32), $k_{ij}$, in the Laplace transformed, s space:
\begin{widetext}

\begin{eqnarray*}
k_{ff} &=&e^{\frac{1}{2}\tau (-2b_{1}+\Phi _{f}+\Phi
_{g}-R)}(b_{1}(a_{1}\Theta _{f}\Theta _{g}(-\Phi _{f}-\Phi
_{g}+R)+a_{1}e^{\tau R}\Theta _{f}\Theta _{g}(\Phi _{f}+\Phi _{g}+R) \\
&&+b_{1}^{2}e^{b_{1}^{\tau }}((1-e^{\tau R})(\Phi _{f}-\Phi _{g})-R-e^{\tau
R}R)+e^{\tau (b_{1}+R)}(\Phi _{f}^{3}+\Phi _{f}^{2}(-3\Phi _{g}+R)+ \\
&&(\Phi _{g}^2-(-4+a_{1})\Theta _{f}\Theta _{g})(-\Phi _{g}+R)+\Phi _{f}(3\Phi
_{g}^{2}+(4+a_{1})\Theta _{f}\Theta _{g}-2\Phi _{g}R))+ \\
&&e^{b_{1}\tau }(-\Phi _{f}^{3}+(\Phi _{g}^{2}-(-4+a_{1})\Theta _{f}\Theta
_{g})(\Phi _{g}+R)+\Phi _{f}^{2}(3\Phi _{g}+R)-\Phi _{f}(3\Phi _{g}^{2}+ \\
&&(4+a_{1})\Theta _{f}\Theta _{g}+2\Phi _{g}R)))+4a_{1}e^{\frac{1}{2}\tau
(b_{1}+R)}(R(-b_{1}^{2}\Phi _{f}+\Phi _{f}(\Phi _{f}-\Phi _{g})^{2}+ \\
&&(3\Phi _{f}-\Phi _{g})\Theta _{f}\Theta _{g}\cosh (\frac{1}{2}\tau R)\sinh
(\frac{b_{1}\tau }{2})+(b_{1}^{2}\Theta _{f}\Theta _{g}\cosh (\frac{%
b_{1}\tau }{2})-(b_{1}^{2}- \\
&&(\Phi _{f}-\Phi _{g})^{2}-4\Theta _{f}\Theta _{g})(\Phi _{f}^{2}-\Phi
_{f}\Phi _{g}+\Theta _{f}\Theta _{g}) \\
&&\sinh (\frac{b_{1}\tau }{2}))\sinh (\frac{1}{2}\tau
R))))/(2b_{1}R(-b_{1}^{2}+R^{2}))
\end{eqnarray*}

\begin{eqnarray*}
k_{gf} &=&(e^{\frac{1}{2}\tau (-2b_{1}+\Phi _{f}+\Phi _{g}-R)}(b_{1}\Theta
_{g}(-2b_{1}^{2}e^{b_{1}\tau }(-1+e^{\tau R})+e^{\tau (b_{1}+R)}(2\Phi
_{f}^{2}-(4+a_{1}) \\
&&\Phi _{f}\Phi _{g}+(2+a_{1})\Phi _{g}^{2}+2(4+a_{1})\Theta _{f}\Theta
_{g}-a_{1}\Phi _{g}R)-e^{b_{1}\tau }(2\Phi _{f}^{2}-(4+a_{1}) \\
&&\Phi _{f}\Phi _{g}+(2+a_{1})\Phi _{g}^{2}+2(4+a_{1})\Theta _{f}\Theta
_{g}+a_{1}\Phi _{g}R)+a_{1}(-2\Theta _{f}\Theta _{g}+\Phi _{g} \\
&&(\Phi _{f}-\Phi _{g}+R))+a_{1}e^{\tau R}(2\Theta _{f}\Theta _{g}+\Phi
_{g}(-\Phi _{f}+\Phi _{g}+R)))+4a_{1}e^{\frac{1}{2}\tau (b_{1}+R)} \\
&&\Theta _{g}((-b_{1}^{2}+\Phi _{f}^{2}-\Phi _{f}\Phi _{g}+2\Theta
_{f}\Theta _{g})R\cosh (\frac{1}{2}\tau R)\sinh (\frac{b_{1}\tau }{2}%
)+(b_{1}^{2}\Phi _{g}\cosh (\frac{b_{1}\tau }{2}) \\
&&+\Phi _{f}(-b_{1}^{2}+R^{2})\sinh (\frac{b_{1}\tau }{2})) \\
&&\sinh (\frac{1}{2}\tau R))))/(2b_{1}R(-b_{1}^{2}+R^{2}))
\end{eqnarray*}

\begin{eqnarray*}
k_{fg} &=&(2e^{\frac{1}{2}(-b_{1}+\Phi _{f}+\Phi _{g})\tau }\Theta
_{f}(-a_{1}((b_{1}+\Phi _{f}-\Phi _{g})(b_{1}+\Phi _{g})-2\Theta _{f}\Theta
_{g})R\cosh (\frac{1}{2}\tau R) \\
&&\sinh (\frac{b_{1}\tau }{2})+(b_{1}(-(b_{1}+\Phi _{f}-\Phi
_{g})(b_{1}-(1+a_{1})\Phi _{f}+\Phi _{g})+2(2+a_{1})\Theta _{f}\Theta _{g})
\\
&&\cosh (\frac{b_{1}\tau }{2})-(b_{1}+a_{1}\Phi _{g})(b_{1}^{2}-(\Phi
_{f}-\Phi _{g})^{2}-4\Theta _{f}\Theta _{g})\sinh (\frac{b_{1}\tau }{2})) \\
&&\sinh (\frac{1}{2}\tau R)))/(b_{1}R(-b_{1}^{2}+R^{2}))
\end{eqnarray*}

\begin{eqnarray*}
k_{gg} &=&(e^{\frac{1}{2}\tau (-2b_{1}+\Phi _{f}+\Phi
_{g}-R)}(b_{1}(a_{1}\Theta _{f}\Theta _{g}(-\Phi _{f}-\Phi
_{g}+R)+a_{1}e^{\tau R}\Theta _{f}\Theta _{g} \\
&&(\Phi _{f}+\Phi _{g}+R)+b_{1}^{2}e^{b_{1}\tau }((-1+e^{\tau R})\Phi
_{f}+\Phi _{g}-R-e^{\tau R}(\Phi _{g}+R))+ \\
&&e^{b_{1}\tau }(\Phi _{f}^{3}-\Phi _{g}^{3}-(4+a_{1})\Phi _{g}\Theta
_{f}\Theta _{g}+\Phi _{g}^{2}R-(-4+a_{1})\Theta _{f}\Theta _{g}R+ \\
&&\Phi _{f}^{2}(-3\Phi _{g}+R)+\Phi _{f}(3\Phi _{g}^{2}-(-4+a_{1})\Theta
_{f}\Theta _{g}-2\Phi _{g}R))+e^{\tau (b_{1}+R)} \\
&&(-\Phi _{f}^{3}+\Phi _{g}^{3}+(4+a_{1})\Phi _{g}\Theta _{f}\Theta _{g}+\Phi
_{g}^{2}R-(-4+a_{1})\Theta _{f}\Theta _{g}R+\Phi _{f}^{2} \\
&&(3\Phi _{g}+R)+\Phi _{f}(-3\Phi _{g}^{2}+(-4+a_{1})\Theta _{f}\Theta
_{g}-2\Phi _{g}R)))+ \\
&&4a_{1}e^{\frac{1}{2}\tau (b_{1}+R)}(R(-b_{1}^{2}\Phi _{g}+(\Phi _{f}-\Phi
_{g})^{2}\Phi _{g}-(\Phi _{f}-3\Phi _{g})\Theta _{f}\Theta _{g}) \\
&&\cosh (\frac{1}{2}\tau R)\sinh (\frac{b_{1}\tau }{2})+(b_{1}^{2}\Theta
_{f}\Theta _{g}\cosh (\frac{b_{1}\tau }{2})+(b_{1}^{2}-(\Phi _{f}-\Phi
_{g})^{2}- \\
&&4\Theta _{f}\Theta _{g})((\Phi _{f}-\Phi _{g})\Phi _{g}-\Theta _{f}\Theta
_{g})\sinh (\frac{b_{1}\tau }{2})) \\
&&\sinh (\frac{1}{2}\tau R))))/(2b_{1}R(-b_{1}^{2}+R^{2}))
\end{eqnarray*}

\[
R=\sqrt{(\Phi _{f}-\Phi _{g})^{2}+4\Theta _{f}\Theta _{g}}
\]

\end{widetext}

\end{document}